\documentclass[aps,twocolumn,10pt,superscriptaddress,prb]{revtex4-1}

\usepackage{amsmath,amssymb,amsfonts}
\usepackage{graphicx}
\usepackage[colorlinks=True,linkcolor=red,citecolor=blue,urlcolor=blue]{hyperref}
\usepackage{xcolor}
\usepackage{chngcntr}

\newcommand{\vk}{\mathbf{k}}
\newcommand{\half}{\frac{1}{2}}
\def\be{\begin{equation}}
\def\ee{\end{equation}}
\def\t#1{\textrm{#1}}

\begin{document}

\title{Chiral Optical Response of Multifold Fermions}
\author{Felix Flicker}
\affiliation{Rudolph Peierls Centre for Theoretical Physics, University of Oxford, Department of Physics, Clarendon Laboratory, Parks Road, Oxford, OX1 3PU, United Kingdom}
\affiliation{Department of Physics, University of California, Berkeley, CA 94720, USA}
\author{Fernando de Juan}
\affiliation{Rudolph Peierls Centre for Theoretical Physics, University of Oxford, Department of Physics, Clarendon Laboratory, Parks Road, Oxford, OX1 3PU, United Kingdom}
\author{Barry Bradlyn}
\affiliation{Princeton Center for Theoretical Science, Princeton University, Princeton, New Jersey 08544, USA}
\author{Takahiro Morimoto}
\affiliation{Department of Physics, University of California, Berkeley, CA 94720, USA}
\author{Maia G. Vergniory}
\affiliation{Donostia International Physics Center, 20018 Donostia-San Sebastian, Spain}
\affiliation{IKERBASQUE, Basque Foundation for Science, Maria Diaz de Haro 3, 48013 Bilbao, Spain}
\author{Adolfo G. Grushin}
\affiliation{Institut N\'eel, CNRS and Universit\'e Grenoble Alpes, F-38042 Grenoble, France}
\date{25 June 2018}

\begin{abstract}
Multifold fermions are generalizations of two-fold degenerate Weyl fermions with three-, four-, six- or eight-fold degeneracies protected by crystal symmetries, of which only the last type is necessarily non-chiral. Their low energy degrees of freedom can be described as emergent relativistic particles not present in the Standard Model of particle physics. We propose a range of experimental probes for multifold fermions in chiral symmetry groups based on the gyrotropic magnetic effect (GME) and the circular photo-galvanic effect (CPGE). We find that, in contrast to Weyl fermions, multifold fermions can have zero Berry curvature yet a finite GME, leading to an enhanced response. The CPGE is quantized and independent of frequency provided that the frequency region at which it is probed defines closed optically-activated momentum surfaces. We confirm the above properties by calculations in symmetry-restricted tight binding models with realistic density functional theory parameters. We identify a range of previously-unidentified ternary compounds able to exhibit chiral multifold fermions of all types (including a range of materials in the families AsBaPt and Gd$_3$Cl$_3$C), and provide specific predictions for the known multifold material RhSi.
\end{abstract}

\maketitle

%
\section{Introduction}\label{sec:introduction}
%

Weyl fermions are chiral, massless spin-$1/2$ particles obeying the Weyl equation~\cite{Weyl29}. Predicted shortly after the formulation of the Dirac equation, Weyl fermions have yet to be discovered as fundamental particles. However, condensed matter analogues of Weyl fermions have been proposed and experimentally realized in so-called Weyl semimetals (WSMs)~\cite{ArmitageEA18,TurnerVishwanath13,HosurQi13,Burkov15}, in which they exist as two-band crossings of linearly-dispersing bands. The crossings points -- known as Weyl nodes -- are perturbatively stable on dimensional grounds\cite{Herring,Weng2015}, a fact which carries with it a topological interpretation: each node carries a monopole of Berry curvature, and hence a gap can only be opened when nodes of opposite charge annihilate. The Berry curvature causes those fermions with crystal momenta close to a node to behave as if they are in the presence of an effective magnetic monopole. These monopoles are responsible for many exotic physical properties of Weyl semimetals, including protected surface Fermi arcs~\cite{WanEA11} connecting the bulk nodes and unconventional magnetoresistance~\cite{NielsenNinomiya,Son13}.

\begin{figure*}  
\includegraphics[width=\linewidth]{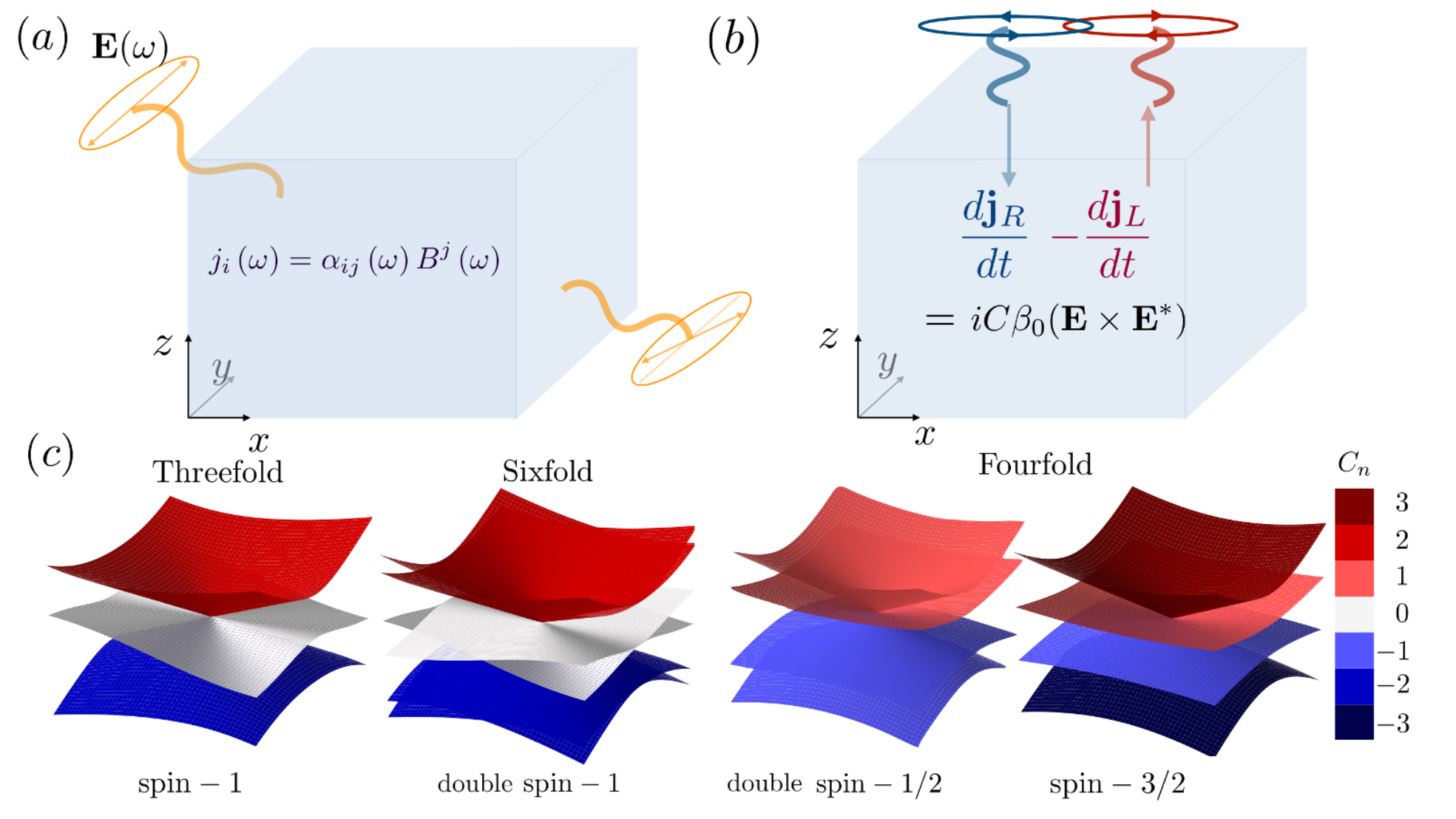}
\caption{\label{fig:introfigure}
\textbf{Chiral optical responses and multifold fermions}. 
Top: Schematics of the physical effects considered in this work and the response functions that describe them. {\bf a)} The gyrotropic magnetic effect (GME), a current response parallel to the direction of an applied low-frequency magnetic field. The GME is the low-frequency limit of natural optical activity, the rotation of the plane of linearly-polarized light upon transmission through a gyrotropic material. {\bf (b)} The circular photogalvanic effect (CPGE) is the photo generation of a current which changes sign under a change in the polarization of the incident light, and which is proportional to the light's intensity. The CPGE is quantized to an integer (given by a combination of Chern numbers)  multiple of $\beta_0 =\frac{\pi e^3}{h^2}$ for chiral multifold fermions.  
{\bf (c)} Schematic dispersion relations for chiral multifold fermions. 
The threefold node harbors effective spin-1 quasiparticles. The Chern numbers of the bands are $C=-2,0,2$ from low to high. The sixfold node is a symmetry-protected doubling of the spin-1 node. The first fourfold node is a symmetry-protected doubling of a standard spin-1/2 Weyl node, and the bands have Chern numbers $C=-1,-1,1,1$ from low to high. The second node realises effective spin-3/2 quasiparticles. The Chern numbers of the bands are $C=-3,-1,1,3$ from low to high. }
\end{figure*}

Weyl excitations exist in Helium-3\cite{Bevan1997,KV05} and in solid state systems, where they were first predicted~\cite{Weng2015,Huang15} and realized in the TaAs family of materials~\cite{LvEA15,XuEA15B}. The corresponding Weyl nodes and the associated Fermi arcs were identified with angle resolved photoemission spectroscopy (ARPES) and quasiparticle interference (QPI) in surface tunneling measurements~\cite{LvEA15A,LvEA15B,XuEA15A,XuEA15B,XuEA15C,BelopolskiEA16B,BelopolskiEA16A,YangEA15}.

The topological stability of Weyl nodes allows them to be present in any crystal symmetry group, provided that either time-reversal or inversion symmetry (or both) is broken\cite{ArmitageEA18}. In particular, the existence of Weyl nodes does not require a chiral crystal structure (one with only orientation-preserving symmetries). In chiral crystals, however, Weyl nodes of opposite charge need not be coincident in energy. Furthermore, Weyl nodes in chiral crystals may occur at time-reversal invariant momenta (TRIMs\footnote{We apologize for this abuse of grammar.})\cite{manes2012,Berneviglecture,kramersweyl}. Recent work has identified that a wider variety of topologically charged fermions beyond the Weyl paradigm can appear in condensed matter systems, unconstrained by Lorentz invariance and the spin-statistics connection~\cite{Fang2012,Tsirkin2017,type2weyl,WiederDD,BradlynEA17}. In the chiral space groups, there exist three-, four-, and sixfold band degeneracies protected by crystal symmetries, which we refer to collectively as \emph{multifold fermions}\cite{manes2012,BradlynEA17,ChangEA17,TangEA17,Bouhon2017}. In this work, we investigate how multifold fermions may be probed in electrical and optical experiments. We will focus on the unique interplay between topological charge and magneto-electric and polarization-dependent response. 

As with bands emanating from a Weyl node, bands meeting in a multifold degeneracy in a chiral group can be assigned (generically non-zero) Chern numbers, defined by the flux of the Berry curvature through a closed surface in momentum space. Certain non-chiral space groups are able to exhibit eight-fold and alternative sixfold crossings, but the bands cannot be assigned Chern numbers, so are not topologically charged in the sense considered here~\cite{WiederDD,BradlynEA17}. We do not consider these structures in this paper, although we occasionally refer to \emph{chiral} multifold fermions to emphasise that these cases are excluded. In Fig.~\ref{fig:introfigure} we show the four possible chiral multifold nodes and the associated Chern numbers of the bands. In each case the corresponding low-energy effective Hamiltonian at the node can be written as a generalization of the Weyl form $\hat{H}=\mathbf{k}\cdot \mathbf{S}$, and in many cases the matrices $S$ additionally take the form of higher-spin representations of $SU(2)$. Multifold fermions, predicted to occur for example in the chiral semi-metal RhSi, feature many Fermi arcs with intricate connectivity, making ARPES experiments challenging~\cite{TangEA17,ChangEA17}. Bulk transport and optical probes could provide alternative tests of their existence, as in Weyl semimetals~\cite{BurkovBalents11,HosurEA12,OrlitaEA14,ChenEA15,SushkovEA15}, but until now have been unexplored for multifold fermions.

In this work, we study two bulk transport responses that display specific, unique features of the chiral multifold fermions: the gyrotropic magnetic effect (GME) and the circular photogalvanic effect (CPGE).
The GME is the low-frequency limit of `natural optical activity', the rotation of the plane of linearly-polarized light upon transmission through an inversion-broken (gyrotropic) material~\cite{Arago11,Nye,LLP} [see Fig.~\eqref{fig:introfigure} (a)].
It has been extensively revisited recently in the context of Weyl semimetals~\cite{GoswamiEA15,MaPesin15,ZhongEA16,MaPesin17,Rou17} due to its connection to the chiral magnetic effect~\cite{Alekseev98,FKW08,Kharzeev09}, which is defined as the existence of a current flowing parallel to a magnetic field $\mathbf{B}$~\cite{ZWB12,ZB12,G12,GT13b,ChenEA13,Vazifeh13,GT13,ChangEA15A,ChangEA15B}. Such a response can only occur under static fields with a non-equilibrium band population~\cite{Vazifeh13,Landsteiner14}, or as a low frequency response to dynamical electromagnetic fields. The latter connects to the GME expressed using Faraday's law $\mathbf{B} = \mathbf{q}\times\mathbf{E}/\omega$ as the electric current response generated by a low frequency oscillating magnetic field.

The GME tensor measures the magnetic moment of bands at the Fermi level~\cite{MaPesin15}. In two band models, such as those around Weyl nodes, the orbital moment is accidentally proportional to the Berry curvature. In this work we find that for multifold fermions the bands' GME responses and Chern numbers not only fail to be proportional, but in many cases they form reversed hierarchies, in which bands with the largest orbital moment have the smallest (even zero) Berry curvature (integrating both quantities over some surface enclosing the node). We further identify a signature of threefold fermions in the form of a discontinuous derivative in the GME as a function of chemical potential, and note that the magnitude of the GME response in these systems is generically enhanced relative to Weyl semimetals and other chiral materials.


The second effect we study is the CPGE [see Fig.~\eqref{fig:introfigure} (b)], which contributes to the nonlinear response, second order in electric field, describing the (DC) current that flows in response to an incident light pulse. In particular, the CPGE is the contribution to the response that switches sign when reversing the sense of the polarization of light~\cite{SipeShkrebtii,NastosSipe,AversaSipe,Deyo09}.
Second order nonlinear effects require the breaking of inversion symmetry, a  condition that permits the existence of topologically charged Weyl points. This observation~\cite{MooreOrenstein10,MorimotoEA16} naturally places Weyl semimetal materials under the spotlight for the search of novel and enhanced photovoltaic phenomena~\cite{Chan17,Taguchi16,deJuanEA17,Konig17,Yang17,Golub18,Zhang18,CLB16,Cortijo16,KKM18}. Indeed, experimentally, nonlinear responses are significantly enhanced~\cite{Ma17,Sun17,Burch17,Rappe18,Patankar18,MaGu18}, yet the underlying microscopic mechanisms are yet to be thoroughly understood.

Along with these observations recent work has predicted that the trace of the CPGE tensor is quantized in units of $e^3/h^2$ for a range of frequencies in Weyl semimetals, as it is directly determined by the monopole charge of the nodes~\cite{deJuanEA17}. It was shown that this quantization is exact only in two band models: the presence of additional bands introduces non-universal corrections that scale with frequency $\omega$, and their energy separation to the Weyl point $\Delta E$, as $\omega^2/\Delta E^2$. The presence of these corrections severely restricts the search for potential candidates to present the effect, since the sub-set of the two Weyl bands must be significantly far away in energy from all other states in the spectrum. Naively, one might think that these corrections exclude the possibility of a quantized CPGE trace in multifold fermions, contrary to what was suggested in Ref.~\onlinecite{ChangEA17}. Since multifold fermions are protected crossings of several bands, the correction $\omega^2/\Delta E^2$ would diverge since $\Delta E^2\to 0$ for any frequency. 

In this work we prove that this simplistic reasoning is incorrect and that the CPGE trace is in fact quantized to different integer values related to the monopole charges of the bands, for any chiral multifold fermion in a range of frequencies, provided some conditions are met, which we derive. 
Together with the unique GME responses, we therefore outline a range of unique signatures of multifold fermions applicable to a wide variety of bulk transport and optical set-ups. Note that both effects are the response of the electric current to a pseudovector ($\mathbf{B}$ in the GME, $\mathbf{E}\times \mathbf{E}^*$ in the CPGE). As a result, the traces of the GME and CPGE tensors vanish in the presence of any orientation-reversing operation, and can only be non-zero in the chiral (or enantiomorphic) point groups. We therefore study, beyond low energy models, realistic tight binding Hamiltonians, taking into account the effect of the (often overlooked) orbital embeddings. Using first principles calculations, we predict a range of previously unexplored materials (in the families Gd$_3$Cl$_3$C and AsBaPt) as being able to demonstrate the phenomena outlined in this work. This adds to the previously-identified multifold material family containing RhSi~\cite{ChangEA17,TangEA17,BradlynEA17}.

This paper proceeds as follows. In Section~\ref{sec:background} we review some relevant information on multifold fermions, as well as the GME and CPGE. In Section~\ref{sec:GME_analytics} we provide analytical calculations of the GME responses of each type of multifold node. In Section~\ref{sec:CPGE_analytics} we provide both analytical and numerical calculations of the nodes' CPGE responses. In Section~\ref{sec:realistic_material_hamiltonians} we provide numerical results for the GME in realistic band structures in groups 198 and 199; between them these models feature all possible multifold node types. We additionally provide specific predictions for transport and optical experiments in RhSi, a material in space group 198. In Section~\ref{sec:abinitio} we present the results of \emph{ab initio} calculations which result in materials where clear multifold nodes lie close to the Fermi level. Finally, in Section~\ref{sec:conclusions} we provide concluding remarks. 

Technical details are left to the appendices, which are as follows. Appendix~\ref{appendix:GME} features calculational details of the GME. Appendix~\ref{appendix:symmetries} clarifies which phenomena can occur in which inversion-broken space groups. Appendices~\ref{appendix:H3f} and~\ref{app:4f} provide formulae for the frequency and energy scales of three-fold and four-fold fermions respectively, referred to in the CPGE plots in the main text. Appendix~\ref{appendix:ab_initio} provides details of the \emph{ab initio} methods employed in the search for new materials. Appendix~\ref{appendix:tight_binding} provides details of the construction of the tight-binding models used throughout the paper as well as the models themselves. Finally, Appendix~\ref{appendix:kdphams} demonstrates the decoupling of the doubled fermion structures, along with additional details of the $\mathbf{k}\cdot\mathbf{p}$ Hamiltonians used in the text.
%
\section{Background}\label{sec:background}
%

In this section we provide some relevant background to multifold fermions, the gyrotropic magnetic effect (GME), and the circular photogalvanic effect (CPGE), as well as a number of original explanations and clarifications.

\subsection{Multifold Fermions}\label{sec:multifold_intro}

\begin{table*}
\begin{tabular}{c|c|c|c|c}
 node & $C_n$ & $D_n$ & No SO & SO \\
 \hline 
Threefold (spin-$1$) & $-2,0,2$ & $1,2,1$ & $195-199$, $207-214$ & $199,214$ \\
Sixfold (doubled spin-$1$) & $\left(-2,0,2\right)\times 2$ & $\left(1,2,1\right)\times 2$ & -- & $198,212,213$ \\
Fourfold (spin-$3/2$) & $-3,-1,1,3$ & $\frac{3}{2},\frac{7}{2},\frac{7}{2},\frac{3}{2}$ & -- & $195-199$ , $207-214$ \\
Fourfold (doubled spin-$1/2$) & $\left(-1,1\right)\times 2$ & $\left(1,1\right)\times 2$ & $19,92,96,198,212,213$  & $18,19,90,92,94,96,198,212,213$
\end{tabular}
\caption{\label{table:nodes}
{\bf Multifold fermions in the $65$ chiral space groups}. The names in parentheses indicate the corresponding effective theories appearing for special parameter choices. $C_n$ indicates the Chern number of the bands, and $D_n$ indicates the GME prefactor (see Section~\ref{sec:GME_intro}). The final two columns list space groups featuring each node type, both with and without spin-orbit (SO) coupling.
}
\end{table*}

In the 65 chiral space groups, isolated point degeneracies at high-symmetry points in the Brillouin zone are generically monopole sources of Berry curvature. A Chern number can be defined by integrating the flux of Berry curvature of each band through a closed surface enclosing the node. The simplest known case is that of a Weyl node at a time-reversal invariant momentum (TRIM)\cite{Berneviglecture,kramersweyl} in a spin-orbit coupled material, a twofold Kramers degeneracy where bands have Chern numbers $C=\pm 1$. The presence of extra symmetries can lead to the protection of nonlinear twofold~\cite{Fang2012,Tsirkin2017} or linear higher-fold degeneracies with higher Chern numbers. The latter are the focus of this work, which we refer to as multifold fermions. In systems with negligible spin-orbit coupling (which we can regard as a spinless system since spin degeneracy is trivial), protected degeneracies can also be found, which in general occur in different space groups and high symmetry points compared to the spin-orbit coupled case. An exhaustive enumeration considering both spinless and spinful cases reveals that multifold fermions in the chiral groups come only in four types, which are schematically shown in Fig.~\ref{fig:introfigure} and summarized in Table \ref{table:nodes}. 

The first type can occur at a TRIM at the corners of the BZ in the presence of a twofold screw axis, which, when combined with time-reversal symmetry, can enforce a doubling of the usual twofold degeneracy, in either spin-orbit or spin-orbit free systems. The resulting fourfold crossing can be seen as a double spin-1/2 fermion. Without spin-orbit coupling, a double spin-1/2 fermion can occur at the R point in space groups 19, 198, 212, and 213, and at the A point in space groups 92 and 96. With spin-orbit coupling, double spin-1/2 fermions occur at the S and R points in space groups 18 and 19; the M and A points in space groups 90, 92, 94, and 96; and the M point in space groups 198, 212, and 213. This is summarized in Table \ref{table:nodes}. These data were extracted from Refs.~\onlinecite{Cracknell,Bilbao1,Bilbao2,Bilbao3,Elcoro2017}. Taking space group (SG) 90 at the M (or A) point as a representative example, the Hamiltonian can be written as (see Appendix~\ref{appendix:kdphams} for further details) 
\begin{equation}
H_{90}(\mathbf{k})=\left(\begin{array}{cc}
H_W(\mathbf{k},b) & 0 \\
0 & -H_W^*(\mathbf{k},-b) 
\end{array}\right)
\label{eq:H90}
\end{equation}
where 
\begin{equation}\label{eq:WeylH}
H_W(\mathbf{k},b)=\hbar v_F\left(\begin{array}{cc}
a k_z & c k_-+i b k_+ \\
-i b k_- +c k_+ & -a k_z 
\end{array}\right).
\end{equation}
Here, $a$, $b$, and $c$ are real numbers. Note that bands remain doubly degenerate along the lines $k_x=k_y=0$ (similar decoupling arguments hold for the other space groups which feature double spin-1/2 fermions with spin-orbit coupling, although there are more degrees of freedom in those cases. See Appendix~\ref{appendix:kdphams} for further details. Together these $\mathbf{k}\cdot\mathbf{p}$ Hamiltonians describe all double spin-1/2 fermions.). A double spin-1/2 fermion can also occur in the spinless case as predicted in Ref.~\onlinecite{manes2012}, where the spinless Hamiltonian can be written in the same way with $c=a$ and $b=0$. In these cases bands are doubly degenerate at every point. Since the Berry curvature texture of the Weyl Hamiltonian in Eq.~\eqref{eq:WeylH} is analogous  to that of a spin-$1/2$ in a magnetic field, we may also call this a double spin-$1/2$ fermion. While specific double spin-1/2 fermions in spin-orbit coupled systems have appeared before~\cite{Bouhon2017,kramersweyl}, the full classification and $\mathbf{k}\cdot\mathbf{p}$ theory is a new result of this work.

The rest of the more complicated multifold fermions can only be found in the cubic space groups, and were previously catalogued in a combination of Refs.~\onlinecite{manes2012},\onlinecite{BradlynEA17}, and \onlinecite{ChangEA17}. First, threefold degeneracies can occur at TRIM points in symmorphic groups without spin-orbit coupling, and at non-TRIM points in nonsymmorphic space groups with spin-orbit coupling. These threefold fermions have a Berry curvature texture that is homotopic to that of a spin-$1$ moment in a magnetic field\cite{Berry84}. The most general $\mathbf{k}\cdot\mathbf{p}$ Hamiltonian near these threefold degeneracies takes the form 
\begin{align}\label{eq:H3f}
H_{3f}\left(\phi,\boldsymbol{k}\right)=\hbar v_F \left(\begin{array}{ccc}
0 & e^{i\phi}k_{x} & e^{-i\phi}k_{y}\\
e^{-i\phi}k_{x} & 0 & e^{i\phi}k_{z}\\
e^{i\phi}k_{y} & e^{-i\phi}k_{z} & 0
\end{array}\right)
\end{align}
where the value of the parameter $\phi$ is material dependent in general. In the absence of spin-orbit coupling, time-reversal symmetry restricts $\phi=\pi/2$~\cite{manes2012}, since the threefold degeneracy occurs at a TRIM in these cases. For $\phi=\pi/2 \mod \pi/3$, the threefold Hamiltonian takes the form $H_{3f} = \mathbf{k} \cdot \mathbf{S}$, where the matrices $\mathbf{S}$ form a spin-1 representation of  $SU(2)$. Thus at these special values of $\phi$ the linearized Hamiltonian has full rotational symmetry. Without spin-orbit coupling, these chiral threefold degeneracies can be found at the $\Gamma$ point in space groups $195$--$199$ and $207$--$214$, the R point in space groups $195,\,207$, and $208$, the H point in space groups $199,\,211$ and $214$, and the P point(s) in space groups $197$ and $211$. With spin-orbit coupling, threefold degeneracies can be found in space groups $199$ and $214$ at the P and $-$P points.

Additionally, fourfold degeneracies can be found in the spin-orbit coupled case, corresponding to the restriction of the spin-$3/2$ representation of $SU(2)$ onto its tetrahedral\cite{ChangEA17,TangEA17} or octohedral\cite{BradlynEA17} subgroup (in the tetrahedral case, time-reversal symmetry is required). These fourfold degeneracies have a Berry curvature texture that is homotopic to that of a spin-$3/2$ moment in a magnetic field. The most general $\mathbf{k}\cdot\mathbf{p}$ Hamiltonian for the octahedral case is:
\begin{align}\label{eq:4f}
H_{4f}\!&=\!\!
\begin{pmatrix}
a k_z&0  & -\frac{a+3b}{4}  k_+ & \frac{\sqrt{3}(a-b)}{4} k_-\\
0 &b k_z &  \frac{\sqrt{3}(a-b)}{4} k_- & -\frac{3a+b}{4}k_+  \\
-\frac{a+3b}{4}  k_- &\frac{\sqrt{3}(a-b)}{4} k_+ & -a k_z &0\\
\frac{\sqrt{3}(a-b)}{4}k_+ &-\frac{3a+b}{4}k_- &  0& -b k_z
\end{pmatrix}
\end{align}
where $a = \hbar v_F \cos \chi$ and $b = \hbar v_F \sin \chi$, and $\chi$ is again a material-dependent parameter. An extra term, given in Appendix~\ref{appendix:kdphams},  is present in the tetrahedral case\footnote{This corrects an error in the Supplementary Material of Ref.~\onlinecite{ChangEA17}.}. 
Contrasting with the double spin-$1/2$ fermions discussed above, bands are generically non-degenerate near a spin-$3/2$ degeneracy, except at special values of $\chi$ where the Chern numbers of bands change~\cite{BradlynEA17}. When $\chi = {\rm arctan}(-3)$ or $\chi = {\rm arctan}(-1/3)$, this Hamiltonian takes the form $H_{4f} = \mathbf{k} \cdot \mathbf{S}$ with $\mathbf{S}$ forming a spin-3/2 representation of $SU(2)$, and $H_{4f}$ recovers full $SU(2)$ invariance. As summarized in Table \ref{table:nodes}, a chiral fourfold crossing can be realized in the groups 195--198 and 207--214 at the $\Gamma$ point, as well as at the R point in SGs 207, 208, and the H point in 211 and 214. The fourfold degeneracy in the octahedral space groups 207--214 can arise from the irreducible spin-$3/2$ representation of the octahedral group, while those in the tetrahedral groups 195--198 arise from the paired $^{1}\bar{F}^{2}\bar{F}$ (co-)representation of the tetrahedral group\cite{Elcoro2017,Cracknell,ChangEA17}, as mentioned above. Throughout this work, an unqualified reference to fourfold fermions will refer to these rather than to double spin-1/2 fermions.

Finally, similarly to the `double spin-$1/2$', a `double spin-$1$' fermion with sixfold degeneracy can be protected by cubic symmetry.  The Hamiltonian of a general sixfold fermion to linear order can be written as (see Appendix~\ref{appendix:kdphams} for further details)
\begin{align}
H_{6f} &=
\left(\begin{array}{cc}
H_{3f}(\tfrac{\pi}{2}-\phi,\boldsymbol{k}) & 0\\
0 &H_{3f}(\tfrac{\pi}{2}+\phi,\boldsymbol{k})\label{eq:H6f}
\end{array}\right).
\end{align}

Sixfold fermions can be found in primitive cubic space groups 198, 212 and 213 at the R point. SG 198 can be pictured as resulting from zone folding of the body-centered space group 199, and similarly 212--213 can be seen as the result of zone-folding the body-centered space group 214. In all cases, the P and $-$P points which host threefold degeneracies are folded into R, which therefore hosts a sixfold crossing. 

In summary, there are four types of multifold fermions in the chiral groups. Since for the double spin-1/2 fermions the response is known from the previous literature on Weyl fermions, in this work we study the chiral optical responses of the other three classes, first with the effective $\mathbf{k}\cdot \mathbf{p}$ models, then with lattice tight-binding models in space groups where they are realized. For concreteness, we choose space group 198 with spin-orbit coupling which features a spin-3/2 fermion at $\Gamma$ and a double spin-1 fermion at R, and space group 199 without spin-orbit coupling which realizes spin-1 fermions at $\Gamma$ and H (note also that 198 features a double spin-$1/2$ fermion at M). Several candidate materials have been predicted for SG 198, including CoSi, RhSi, CoGe, and RhGe~\cite{TangEA17,GellerWood54,BradlynEA17}. In this work we also present a new family of materials in SG 198 and two new materials in SG 214 with spin-1 fermions near the Fermi level. 

\subsection{The Gyrotropic Magnetic Effect}\label{sec:GME_intro}

The Gyrotropic Magnetic Effect (GME) is the electric current response to a low-frequency magnetic field $B^j\left(\omega\right)$:
\begin{align}\label{eq:j_GME}
j_i\left(\omega\right)=\alpha_{ij}\left(\omega\right)B^j\left(\omega\right)
\end{align}
where indices $i,j,k,\ldots$ span cartesian directions. Repeated indices $i,j,k,\ldots$ are summed over throughout the paper unless otherwise stated. The GME can be understood as the low-frequency limit of `natural optical activity', the rotation of the plane of linearly-polarized light upon transmission through an optically active material~\cite{Arago11,Nye,LLP} (schematically shown in Fig.~\ref{fig:introfigure}, top left panel). The GME tensor $\alpha_{ij}$ also characterizes the `inverse GME'\cite{Yoda15,ZhongEA16,SouzaTe,Sahin18,Zhou1,Zhou2,TaikiEA18,BurkovEA18}, where a magnetization $M_i$ is produced in response to an applied electric field (in the presence of time-reversal symmetry):
\begin{align}
M_i\left(\omega\right)=-i\omega^{-1}\alpha_{ji}\left(\omega\right)E^j\left(\omega\right).
\end{align}
Appendix~\ref{appendix:symmetries} provides details of the responses possible in optically active point groups and places them in the context of responses from general inversion-broken media. In particular, it should be noted that natural optical activity can only be measured in transmission but not in reflection~\cite{HosurEA13,Fried14}. 

The GME tensor receives both inter- and intra-band contributions~\cite{ZhongEA16}. At low frequencies, the intraband contribution dominates and is given by the Fermi surface integral
\begin{align}\label{eq:alpha}
\alpha_{ij}\left(\omega\right)=\frac{i\omega\tau}{i\omega\tau-1}\frac{e}{\left(2\pi\right)^2 h}\sum_{n,a}\int_{\mathrm{FS}}\text{d}S_a\hat{v}_{Fi}^{n}m_{j}^{n}
\end{align}
where $a$ labels the different Fermi surface pockets, $n$ is the band index, $\hat{v}_{Fi}=v_{Fi}/|v_F|$ is $i$-th component of the normalized Fermi velocity, and $\tau$ is the scattering time~\cite{ZhongEA16,MaPesin15}, which in this work is taken to be $\tau\rightarrow\infty$. In the opposite limit, $\omega\ll 1/\tau$, only a dissipative inverse GME exists~\cite{ZhongEA16}.

The quantity $m^n_j\left(\mathbf{k}\right)=m^{n}_{\textrm{orb},i}\left(\mathbf{k}\right)+S^{n}_i\left(\mathbf{k}\right)$ is the magnetic moment of band $n$ at wavevector $\mathbf{k}$~\cite{ZhongEA16,ChangNiu95}, whose orbital part is given by
\begin{align}\label{eq:m}
m^{n}_{\textrm{orb},i}\left(\mathbf{k}\right)=\frac{i}{2}\frac{e}{\hbar}\epsilon_{ijk}\sum_{n'\ne n}\frac{\langle n|\left(\partial^{j}H_{\mathbf{k}}\right)|n'\rangle\langle n'|\left(\partial^{k}H_{\mathbf{k}}\right)|n\rangle}{\epsilon_{n'}-\epsilon_{n}}
\end{align}
where $\partial^i=\partial/\partial k_i$, and $\epsilon_{n\mathbf{k}}$ and $|n\rangle$ are the energies and wavefunctions of the Bloch Hamiltonian $H_{\mathbf{k}}|n\rangle = \epsilon_{n\mathbf{k}} |n\rangle$. The spin contribution to the magnetic moment is given by
\begin{align}
S^{n}_i=-\frac{eg_s\hbar}{4m_e}\langle n|\sigma_i | n\rangle
\end{align}
where $e$, $m_e$ and $g\simeq 2$ are the charge, mass, and spin factor of the electron respectively. Note that $\sigma_i$ here refers to the real spin, as opposed to the pseudospin referred to in the $\mathbf{k}\cdot\mathbf{p}$ Hamiltonians of the previous section. For the rest of this work, we will only consider the trace of the GME tensor, which we designate $\alpha = {\rm tr}\left(\alpha_{ij}\right)$. 

\subsection{The Circular Photogalvanic Effect}\label{sec:CPGE_intro}

In the circular photogalvanic effect (CPGE), circularly polarized light incident on a gyrotropic material causes a time-dependent current density which relaxes to a DC current response. This response can be produced by different mechanisms, and in this work we will concentrate on the intrinsic contribution at $\tau\rightarrow\infty$, termed the `injection current', given by 
\begin{equation}\label{eq:injection}
\dfrac{d j_i}{dt} = \beta_{ij}(\omega) \left[\mathbf{E}(\omega)\times \mathbf{E}^{*}(\omega)\right]^{j},
\end{equation}
where $\mathbf{E}(\omega)=\mathbf{E}^{*}(-\omega)$ is the electric field (the star indicates complex conjugation). This current grows linearly in time for $t\ll\tau$. The CPGE tensor $\beta_{ij}$ can be written in general as~\cite{SipeShkrebtii}:
\begin{eqnarray}\label{eq:nu}
\beta_{ij}(\omega)\!=\!\!\dfrac{\pi e^3}{\hbar V}\epsilon_{jkl}  \sum_{\vk,n,m} f^{\vk}_{nm} \Delta^i_{\vk,mn} r^k_{\vk,nm} r^l_{\vk,mn} \delta(E_{\vk,mn}-\hbar\omega)\nonumber\\
\end{eqnarray}
where $V$ is the sample volume, $E_{\vk,nm}=E_{\vk,n}-E_{\vk,m}$ and $f^{\vk}_{nm}=f^{\vk}_{n}-f^{\vk}_{m}$ are the differences between band energies and Fermi-Dirac distributions respectively, 
$\mathbf{r}_{\vk,nm} = i \left<n|\partial_{\vk}|m\right>$ is the cross gap Berry connection, and $\Delta^i_{\vk,nm} = \partial_{k_i}E_{\vk,nm}/\hbar$. For the rest of this work, we will only focus on the trace of the CPGE tensor, which we label as $\beta = {\rm tr}\left(\beta_{ij}\right)$. 

%
\section{Gyrotropic Magnetic Effect: Results for Low-Energy Effective Models}\label{sec:GME_analytics}
%

We first consider the GME produced by all types of chiral multifold fermions with effective low energy models. As we consider crossings with more than two bands, the Berry curvature (defining the topology of the node) and the orbital moment (responsible for GME) are independent, as shown in Table \ref{table:nodes}. To emphasize this distinction, in this section we compute the GME for the multifold fermions in the limit where their Hamiltonians have full rotational invariance. 

\subsection{GME of Threefold Fermions}

The generic low energy Hamiltonian of a threefold fermion is given in Eq.~\eqref{eq:H3f}. The Hamiltonian can be written $H=\hbar v_F k_i S^i$, and at the special point $\phi=\pi/2$ the $S^i$ form a spin-1 representation of $SU(2)$. Close to the node the energies are $E_{1,2,3}\left(\mathbf{k}\right) = \hbar v_F k,\,0,\,-\hbar v_F k$, where $k=\left|\mathbf{k}\right|$ and the bands have Chern numbers $C_{1,2,3}=2,\,0,\,-2$.

The orbital magnetic moment can be calculated using Eq.~\eqref{eq:m}. Assuming a small spherical Fermi surface pocket around the node, the result is found in Appendix~\ref{appendix:GME} to be:
\begin{align}\label{eq:OrbMom3}
m^{n}_{\mathrm{orb},i} = \frac{e}{2} D_{n} v_F \dfrac{k_i}{k^{2}}
\end{align}
with $D_{n}=1,\,2,\,1$ for the three bands respectively. There are three important points to note. First, the value of $D_n$ depends geometrically on the surface taken around the node, unlike the Chern number $C_n$ which is topological. Second, $m^{n}_{\mathrm{orb},i} $ is equal for the two bands in which the Chern number is opposite. Third, the band which has zero Berry curvature and zero Chern number has the largest orbital moment of all the three bands. The last statement shows that, not only is the magnitude of the orbital moment (and thus the GME) not related to the Berry curvature, but the hierarchies between the bands are distinct. 

We proceed to calculating the GME tensor $\alpha_{ij}$ of Eq.~\eqref{eq:alpha}. For the analytic results of this section we neglect the spin contribution to the magnetic moment, returning to it in Section~\ref{sec:realistic_material_hamiltonians}. The effective model of Eq.~\eqref{eq:alpha} at $\phi=\pi/2$ presents the problem that the middle band is completely flat and does not form a Fermi surface at any chemical potential $\mu$. The GME from this band is determined from quadratic corrections to $H$. With the simple term $H^{(2)} = \hbar^2\mathbf{k}^2/2m$, which is always allowed by symmetry, the middle band does form a Fermi surface with the same orbital moment. The resulting tensor is then
\begin{align}\label{eq:alpha_threefold}
\alpha_{ij} = \delta_{ij} \frac{1}{3}\frac{e^2}{h^2} \left\{ \begin{array}{c} D_{1}\left(\mu-\epsilon_{\textrm{node}}\right) \Theta(\mu-\epsilon_{\textrm{node}}) \\ D_2 v_F \sqrt{2m \left(\mu-\epsilon_{\textrm{node}}\right)} \Theta\left(\mu-\epsilon_{\textrm{node}}\right) \\ D_{3}|\mu-\epsilon_{\textrm{node}}| \Theta\left(\epsilon_{\textrm{node}-\mu}\right)\end{array} \right.
\end{align}
with $\Theta(x)$ the Heaviside step function. As a result, the dependence on $\mu$ switches from linear to square root behaviour when crossing the node. This is a clear experimental signature unique to (threefold and sixfold) multifold fermions. The results are plotted in Section~\ref{sec:realistic_material_hamiltonians}. We note that in general other, non-diagonal, quadratic corrections are allowed. Since their phenomenology is material-specific we defer a discussion of these effects to the numerical results from TB models in Section~\ref{sec:realistic_material_hamiltonians}.
\subsection{GME of Fourfold Fermions}

The low energy Hamiltonian of fourfold fermions is given in Eq.~\eqref{eq:4f}. The Hamiltonian can again be written in the form $H = v_F \mathbf{k} \cdot \mathbf{S}$, and at the special points $\chi = {\rm arctan} (-3)$ and $\chi = {\rm arctan} (-1/3)$  the matrices $\mathbf{S}$ form a spin-3/2 representation of $SU(2)$. At these points the Berry curvature and orbital moment can be computed analytically. Ordering the four bands $E_1<E_2<E_3<E_4$, for $\chi = {\rm arctan} (-3)$ they have Chern numbers $C_{1,2,3,4} = 3,\,1,\,-1,\,-3$, and for $\chi = {\rm arctan} (-1/3)$ the signs are reversed. A third option, $C_{1,2,3,4}=3,-1,1,-3$, is present in another range of values, but the $S_i$ matrices do not form an irreducible representation of the spin algebra, and we do not consider this case here~\cite{BradlynEA17}. The orbital magnetic moment again takes the form of Eq.~\eqref{eq:OrbMom3}, with prefactors $D_{1,\,2,\,3,\,4} = 3/2,\,7/2,\,7/2,\,3/2$. The GME tensor features a linear dependence on chemical potential close to the node:
\begin{align}\label{eq:alpha_fourfold}
\alpha_{ij} = \delta_{ij} \frac{1}{3}\frac{e^2}{h^2} \left\{ \begin{array}{c} 
D_{1}\left(\mu-\epsilon_{\textrm{node}}\right) \Theta\left(\mu-\epsilon_{\textrm{node}}\right) \\ 
D_{2}\left(\mu-\epsilon_{\textrm{node}}\right) \Theta\left(\mu-\epsilon_{\textrm{node}}\right) \\ 
D_{3}\left|\mu-\epsilon_{\textrm{node}}\right| \Theta\left(\epsilon_{\textrm{node}}-\mu\right) \\
D_{4}\left|\mu-\epsilon_{\textrm{node}}\right| \Theta\left(\epsilon_{\textrm{node}}-\mu\right)\end{array} \right.
\end{align}
where we have again neglected the spin part of the magnetic moment. The results are plotted in Section~\ref{sec:realistic_material_hamiltonians}; different values of $\chi$ cannot be treated analytically, but such cases occur in realistic material models, and are again considered in that section. 
\subsection{GME of Sixfold Fermions}

As shown in Section~\ref{sec:multifold_intro}, and in particular Eq.~\eqref{eq:H6f}, any sixfold Hamiltonian can be brought into a block-diagonal form in which the blocks are the Hamiltonians of threefold nodes. As a result, the contribution of each threefold band is doubled to make the sixfold case. The result is therefore exactly the same as the threefold case with the response doubled. The values of $D_n$ and $C_n$ are given for all multifold node types in Table \ref{table:nodes}.

%
\section{Circular Photogalvanic Effect: Results for Low-Energy Effective Models}\label{sec:CPGE_analytics}
%

As predicted in Ref.~\onlinecite{deJuanEA17}, the trace of the CPGE tensor is exactly quantized for a two band model of a Weyl semi-metal, and quantized up to power law corrections in the presence of extra bands. In this section we first discuss under what conditions the CPGE remains quantized in multifold fermions. Under very general assumptions, we prove that for any model of a multifold fermion that is linear in momentum, there is always a range of frequencies for which the CPGE tensor remains exactly quantized. 

The full CPGE tensor is given in Eq.~\eqref{eq:nu}. Using the terms defined in that equation, and additionally defining the quantity 
\begin{align}
R^j_{nm} =  \epsilon_{jkl}r^k_{nm} r^l_{mn}
\label{eq:R_nm}
\end{align}
where $n,m$ are not summed over, we can write the CPGE trace as
\begin{align}
\beta(\omega) &=  4\pi^2 \beta_0  \int \frac{d^3k}{(2\pi)^3} \sum_{n,m} f_{nm}\partial_{k_i}E_{nm} R^i_{nm} \delta(\hbar\omega - E_{mn}) \nonumber \\
&\equiv 4\pi^2 \beta_0  \sum_{n,m} \int d \vec S_{nm} \cdot \vec R_{nm}
\end{align}
with $\beta_0 =\tfrac{\pi e^3}{h^2}$, so that for every pair $nm$, this integral computes the flux of the vector $R^i_{nm}$ through a manifold $S_{nm}$ defined by the $k$-points for which exactly one of bands $n$ and $m$ is occupied, and the bands are separated in energy by exactly $\omega$. In spherical coordinates this reads
\begin{align}
\beta(\omega) &= 4\pi^2 \beta_0 \int \frac{d^3k}{(2\pi)^3} \sum_{n,m} f_{nm}\frac{\partial_{k_i}E_{nm}}{|\partial_k E_{nm}|}  R^i_{nm} \delta(k - k_{nm}(\theta,\phi))
\end{align}
where $\partial_{k_i} E_{nm}= \partial_k E_{nm} \hat{k}_i + \tfrac{1}{k} \partial_\theta E_{nm}\hat{\theta}_i +\tfrac{1}{k \sin \theta} \partial_\phi E_{nm} \hat{\phi}_i$. If we further assume a Hamiltonian that is linear in momentum for any multifold fermion, several simplifications occur. First, if there is a frequency range where $S_{nm}$ is a closed surface, $f_{nm}=1$ by definition and the $\delta$ function is trivially integrated. Moreover, the lack of energy scale in a linear model ensures that the integrand is $|\mathbf{k}|$-independent, so the result for a closed surface does not depend on $\mu$ or $\omega$. Explicitly, since $R^i_{nm}$ has dimensions of $k^{-2}$, in the absence of any other scale in the problem one may define  $R^i_{nm} = \tfrac{1}{k^2} \bar{R}^i_{nm}$ where $\bar{R}^i_{nm}$ is dimensionless and $k$-independent. The contribution from a closed $S_{nm}$ is then
\begin{align}
4\pi^2 \beta_0 \int d \vec S_{nm} \cdot \vec R_{nm} &= 4\pi^2 \beta_0 \int \frac{d\Omega}{(2\pi)^3} 
\frac{\partial_{k_i}E_{nm}}{|\partial_k E_{nm}|}\bar{R}^i_{nm}
\end{align}
where $n,m$ are not summed over. Next we show that for any linear model, $R_{nm}^i$ is purely radial. This is because $k_i v_{nm}^i = k_i \left<n| \partial_{k_i}H | m\right> = k_i \left<n| S_i | m\right> = \left<n| H | m\right> =0$, so the off-diagonal velocity operators are orthogonal to $k_i$. The same argument applies to $v_{mn}^i$. Since $\vec R_{nm}= \vec v_{nm} \times \vec v_{mn}/(E_n-E_m)^2$, $\vec R_{nm}$ is perpendicular to the plane spanned by $\mathrm{Re}(v_{nm}^i)$ and $\mathrm{Im}(v_{mn}^i)$, which are both orthogonal to $\vec k$. Thus we must have $\vec R_{nm} = \hat k R_{nm}$. In this case, the angular integral simplifies to 
\begin{align}\label{eq:simp}
4\pi^2 \beta_0 \int d \vec S_{nm} \cdot \vec R_{nm}&= 4\pi^2 \beta_0 \int \frac{d\Omega}{(2\pi)^3} \bar{R}_{nm}
\end{align}
($n,m$ are not summed over) which now makes no reference to the shape of the $S_{nm}$ surface. We can then use the relation between $R^i_{nm}$ and Berry curvature
\begin{equation}\label{eq:Omega}
\Omega^c_{n}  = i  \sum_{m\neq n} R^c_{nm}
\end{equation}
to determine in which conditions quantization is possible for the different multifolds~\cite{deJuanEA17}.

For completeness we first review the result in Ref.~\onlinecite{deJuanEA17} for a single Weyl node with top and bottom bands labeled by $n=1,2$, and Chern numbers $C_n =1,-1$. In this case there is only a single surface $S_{12}$, and when this surface is closed 
\begin{equation}
\beta(\omega) =4\pi^2 \beta_0 \int d \vec S_{12} \cdot \vec R_{12} = -i 4\pi^2 \beta_0 \int d \vec S_{12} \cdot \vec \Omega_1 = i C_1 \beta_0
\end{equation}
so the result is quantized for the twofold case. For a double spin-1/2 fermion Hamiltonian, which can be decoupled into two Weyl blocks as shown in Appendix~\ref{appendix:kdphams}, this result applies to each block separately, and the total result is the sum of the two contributions.

For a threefold fermion, we label top middle and bottom bands $n=1,2,3$, with Chern numbers $C_n = 2,0,-2$. If $S_{12}$ and $S_{13}$ are both closed surfaces but $S_{23}$ is Pauli blocked (\emph{i.e.}, $S_{23}$ is an empty set since both bands are either occupied or unoccupied at the resonant frequency), then using $R^c_{12} = -i \Omega_1^c - R^c_{13}$ yields
\begin{align}\label{eq:3fintegral}
\beta(\omega) &=4\pi^2 \beta_0 \left( \int d \vec S_{12} \cdot \vec R_{12} + \int d \vec S_{13} \cdot \vec R_{13} \right)\nonumber \\
 & = 4\pi^2 \beta_0 \left( -i \int d \vec S_{12} \cdot \vec \Omega_1 \right.
 \nonumber \\
 &\qquad\qquad \left. 
 + \left[-\int d \vec S_{12} +\int d \vec S_{13}\right] \cdot \vec R_{13} \right)\nonumber\\ &= i\beta_0 C_1
\end{align}
where we used the fact from Eq.~\eqref{eq:simp} that the shape of the surfaces does not matter to deduce that the correction in square brackets is zero, and the result is quantized. An analogous result would hold if $S_{23}$ and $S_{13}$ were closed. The same result holds for a sixfold fermion, due to the decoupling in Eq.~\eqref{eq:H6f}. 

For a fourfold fermion, we label bands from top to bottom as $n=1,2,3,4$, with corresponding Chern numbers $C_n = 3,1,-1,-3$. If $S_{13}$, $S_{14}$, $S_{23}$, $S_{24}$ are closed and the rest are blocked then 
\begin{align}
\beta(\omega) &=4\pi^2 \beta_0 \left( \int d \vec S_{13} \cdot \vec R_{13} + \int d \vec S_{14} \cdot \vec R_{14}\right. \nonumber \\
&+\left. \int d \vec S_{23} \cdot \vec R_{23} +\int d \vec S_{24} \cdot \vec R_{24}\right) \nonumber \\
 &= 4\pi^2 \beta_0 \left( -i \int d \vec S_{13} \cdot \vec \Omega_1 -i \int d \vec S_{23} \cdot \vec \Omega_2 \right. \nonumber\\&-\int d \vec S_{13} \cdot (\vec R_{12}+\vec R_{14}) +\int d \vec S_{14} \cdot \vec R_{14} \nonumber \\ &- \int d \vec S_{23} \cdot (\vec R_{21}+\vec R_{24}) + \left.\int d \vec S_{24} \cdot \vec R_{24}\right) \nonumber \\ &= i\beta_0 (C_1+C_2)\label{eq:4fintegral}
\end{align}
where we have used  Eq.~\eqref{eq:simp} again to recover the Chern numbers of bands 1 and 2.

\subsection{CPGE of Threefold Fermions}
\begin{figure*}[t]
\begin{center}
\includegraphics[width=\linewidth]{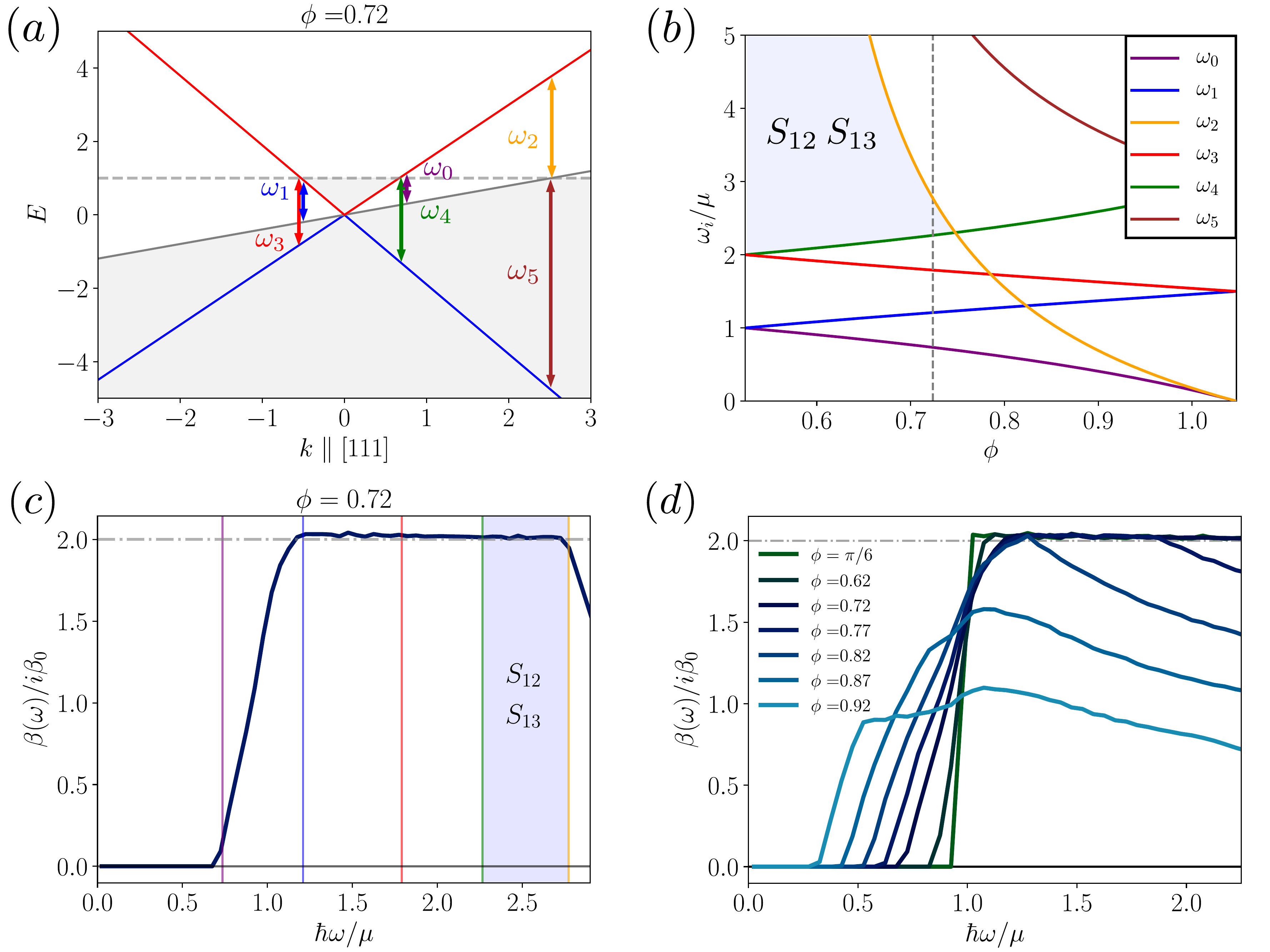}
\caption{\label{fig:spin1_3fold_2}
\textbf{CPGE for an effective 3-fold fermion model:} 
{\bf (a)} A representative band structure of Eq.~\eqref{eq:H3f} for $\phi=\pi/6+0.2\sim 0.72$, with the chemical potential depicted by the dashed horizontal line. The frequencies $\omega_i$ where the different resonant surfaces $S_{nm}$ change from active to inactive and from open to closed are depicted with a consistent color coding with that of (b) and (c). Analytical expressions for the spectrum and for $\omega_i$ can be found in Appendix \ref{appendix:H3f}.
{\bf (b)} Evolution of the different $\omega_i$ as a function of $\phi$. The colored  blue region indicates the region where exact quantization holds due to the fact that $S_{12}$ and $S_{13}$ are closed. In the region $\omega_1<\omega<\omega_2$, $S_{12}$ is closed and a non-quantized CPGE plateau exists.
The vertical gray line corresponds to $\phi=\pi/6+0.2$, the value used to calculate the CPGE in (c). 
{\bf (c)} As in (b) the shaded area denotes the region with exact quantization. Between $\omega_1<\omega<\omega_2$ the plateau is non-universal yet close to the quantized value $2\beta_0$ due to the small magnitude of the corrections. {\bf (d)} CPGE for different values of $\phi$ deviating from the $SU(2)$ invariant case $\phi=\pi/6$. The small deviations (even at $\phi=\pi/6$) from $2\beta_0$ are due to numerical artifacts that decrease with increasing momentum resolution.
}
\end{center}
\end{figure*}

Having determined the conditions under which quantization is possible, we now compute the CPGE explicitly for the threefold fermion with the Hamiltonian in Eq.~\eqref{eq:H3f}, the results of which we present in Fig.~\ref{fig:spin1_3fold_2}. The energies and wavefunctions of this Hamiltonian can be computed analytically for arbitrary $\phi$ and are presented in Appendix \ref{appendix:H3f}. A representative band structure (for $\phi=\pi/6+0.2$) is plotted in Fig.~\ref{fig:spin1_3fold_2} (a). 

To calculate the CPGE we first note that as opposed to the GME calculation, no quadratic corrections are required for a meaningful calculation. The resonant surface of integration for CPGE is defined by $E_{nm}-\omega$ when $n$ is occupied and $m$ unoccupied, and this can be closed at finite $\mu$ in the linear model, despite the presence of open Fermi surfaces. The different frequency ranges where these resonant surfaces become open or closed depend on the parameter $\phi$ (see Fig.~\ref{fig:spin1_3fold_2} (b)) and are bounded by the characteristic energy scales $\omega_i$ with $i\in[0,5]$ depicted in Fig.~\ref{fig:spin1_3fold_2} (a). The manifold $S_{12}$ becomes active for $\omega>\omega_0$ and is closed for $\omega_1 < \omega <\omega_2$ while  $S_{13}$ becomes active for $\omega> \omega_3$ and fully closed for $\omega > \omega_4$. $S_{23}$ becomes active with $\omega> \omega_5$ and is never closed in the linear model.  

It is worth discussing the $\phi = \pi/6$ case first, where the threefold Hamiltonian has full rotational symmetry. In this case the energies are simply $E_n = v_F k, 0, -v_F k$, and there are no open surfaces. $S_{12}$ becomes closed at $\omega_0 =\omega_1=\mu$ and $S_{13}$ becomes closed at $\omega_3 = \omega_4= 2\mu$. However, due to the full rotation symmetry $\vec R_{13} =0$, and according to Eq.~\eqref{eq:3fintegral} we get a fully quantized plateau at $\omega=\mu$ (see Fig.~\ref{fig:spin1_3fold_2} (d)).

When $\phi \neq \pi/6$, $\vec R_{13}$ is finite in general, and according to Eq.~\eqref{eq:3fintegral} we only have a quantized plateau once both $S_{12}$ and $S_{13}$ are closed, which only occurs for $\omega>\omega_4$ indicated by the shaded region in Fig.~\ref{fig:spin1_3fold_2} (b) and (c). When $\omega_1<\omega<\omega_2$, $S_{12}$ is closed and there is a non-quantized plateau that is non-universal. Nevertheless, the contribution from $\vec R_{13}$ is actually quite small, and approximate quantization starts already at $\omega>\omega_1$ as shown in Fig.~\ref{fig:spin1_3fold_2} (c) for the particular case $\phi=\pi/6+0.2$. The plateau is preserved up to $\omega<\omega_2$. If $\phi > \phi_c = {\rm arc tan} (-2+\sqrt{13})/(\sqrt{3}) \approx 0.75 $ the window closes and there is no plateau as seen in  Fig.~\ref{fig:spin1_3fold_2} (d). 

The presence of quadratic terms in the Hamiltonian will have two types of effect. First, the changes in energies will determine the new frequency windows where quantization can be observed. Second, quadratic corrections  will introduce an energy scale $m v_F^2$, with $m$ an effective mass, so that the shapes of the resonant manifolds become important and the cancellations of the correction terms will in general not occur. Quantization will therefore get power law corrections in $\omega/(m v_F^2)$, which can be avoided by measuring at frequencies lower than this scale. It should be noted that power law corrections at higher frequencies are expected anyway in the presence of extra bands beyond those forming the multifold node, which have the same origin as the corrections in the Weyl node case~\cite{deJuanEA17}. 

\subsection{CPGE of Fourfold fermions}

\begin{figure*}[t]
\begin{center}
\includegraphics[width=\linewidth]{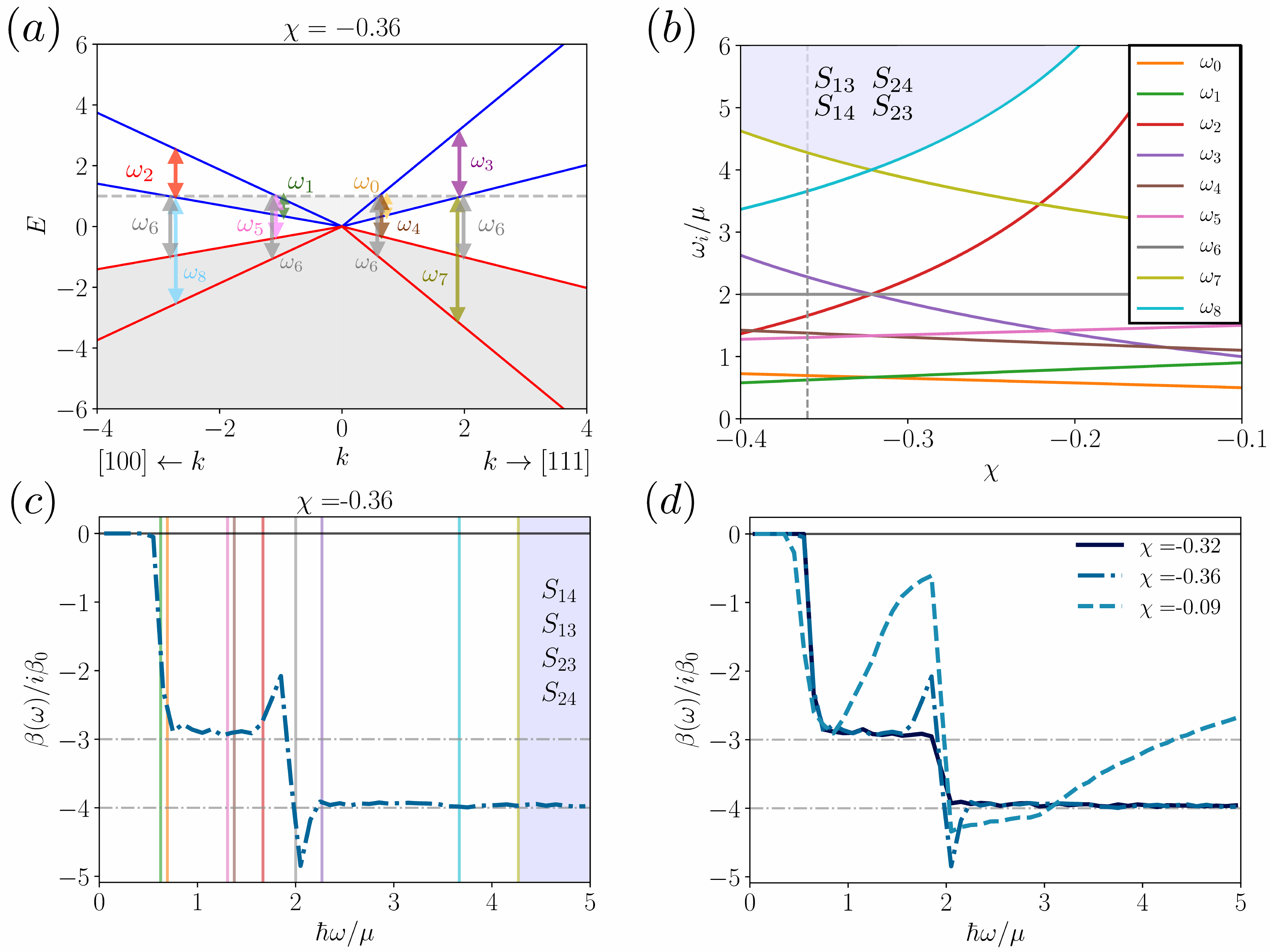}
\caption{\label{fig:4fold}
\textbf{CPGE for an effective 4-fold fermion model:} 
{\bf (a)} A representative band structure of Eq.~\eqref{eq:4f} for $\chi=-0.36$ (see Appendix \ref{app:4f} for analytic expressions for arbitrary $\chi$). The relevant frequency scales $\omega_i$ (see Appendix \ref{app:4f} for  expressions) are depicted with a consistent color coding with that of (b) and (c). The chemical potential is indicated by a dashed horizontal line.
{\bf (b)} Evolution of the different $\omega_i$ as a function of $\chi$ bounding the different types of surfaces $S_{nm}$ (open, closed, inactive or active) of allowed optical transitions. The colored  blue region indicates the region where exact quantization holds due to the fact that $S_{13}$,$S_{14}$,$S_{23}$ and $S_{24}$ are closed. As for the threefold case other regions  have closed surfaces resulting in a non-quantized CPGE plateau.
The vertical gray line corresponds to $\chi=-0.36$, the value used to calculate the CPGE in (c). 
{\bf (c)} The shaded area denotes the region with exact quantization ($4\beta_0$), yet a plateau close to ($3\beta_0$) is seen for $\hbar\omega/\mu\sim 1$. The latter is only exactly quantized at the value of $\chi$ that realizes the spin-3/2 multifold case, $\chi = {\rm arctan} (-3) \approx -0.32$ (solid curve in (d)).
{\bf (d)} The CPGE for different values of $\chi$. 
}
\end{center}
\end{figure*}

Fourfold (spin-3/2) fermions have the low-energy Hamiltonian specified in Eq.~\eqref{eq:4f}. A representative energy spectrum is shown in Fig.~\ref{fig:4fold} (a) (analytic expressions can be found in Appendix \ref{app:4f}).

In analogy with the threefold case, we show in Fig.~\ref{fig:4fold} (a) and (b) the relevant energy scales ($\omega_i$ with $i\in[0,8]$, see Appendix \ref{appendix:H3f} for concrete expressions) where the different resonant surfaces open and close depending on the parameter $-\pi <\chi< 0$ which we recall was defined above as $\chi={\rm arctan}(b/a)$ under Eq.~\eqref{eq:4f}. Taking $\mu>0$ for concreteness, for $-\pi <\chi< {\rm arctan} (-3)$ or ${\rm arctan} (-1/3) <\chi< 0$, $S_{12}$ becomes active for $\omega>\omega_0$, and is closed for $\omega_1 < \omega < \omega_2$. For $\omega_2 <\omega <\omega_3$ it is open again, and becomes blocked at $\omega > \omega_3$. $S_{13}$ becomes active at $\omega = \omega_4$ and closed for any $\omega > \omega_5$. $S_{14}$ and $S_{23}$ are active and already closed for $\omega > \omega_6$. Finally, $S_{24}$ becomes active at $\omega=\omega_7$ and is closed for $\omega > \omega_8$. If ${\rm arctan} (-3) <\chi< {\rm arctan} (-1/3)$, the following frequencies are interchanged: $\omega_0 \leftrightarrow \omega_1$, $\omega_2 \leftrightarrow \omega_3$, $\omega_4 \leftrightarrow \omega_5$, $\omega_7 \leftrightarrow \omega_8$ (see Fig.~\ref{fig:4fold} (b)). 

As with the threefold case, it is instructive to first discuss the case $\chi = {\rm arctan} (-1/3) \approx -0.32$ where full rotational invariance is recovered (solid line in Fig.~\ref{fig:4fold} (d)). In this case the energies are $E_{1}=3\hbar v_F|\mathbf{k}|=3E_2=-3E_3=-E_4$ and optical surfaces are either fully closed or inactive. Due to angular momentum conservation imposed by rotational invariance, $\vec R_{nm}$ can only be nonzero when $|n-m|=1$. Because of this, only two surfaces contribute: $S_{12}$ for $2/3 < \omega/\mu < 2$ and $S_{23}$ for $2 < \omega/\mu$. Furthermore, in this special case $\vec R_{12} = i \Omega_1$ and $\vec R_{23} = i (\Omega_1+\Omega_2)$. This gives rise to two exactly quantized plateaus at $3\beta_0$ and $4\beta_0$ respectively (horizontal gray dashed lines in Fig.~\ref{fig:4fold} (d)). 

In the general case $\chi \neq {\rm arctan} (-1/3)$, several more surfaces contribute. There is still a quantized plateau for $\omega>\omega_7,\omega_8$, when the only active surfaces are $S_{13}$, $S_{14}$, $S_{23}$, $S_{24}$. The region where this happens is shaded blue in Figs.~\ref{fig:4fold} (b) and (c). According to Eq.~\eqref{eq:4fintegral}, this plateau is exactly quantized in the linear model, even without the full rotational invariance. Other plateaus can be found with non-quantized values when other surfaces are closed, but again the deviations from quantization can be accidentally small. As with the threefold case, power law corrections to quantization due to both quadratic terms in the Hamiltonian and the presence of extra bands~\cite{deJuanEA17} are also to be expected. 

\subsection{CPGE of Sixfold Fermions}
As mentioned in Section \ref{sec:multifold_intro}, using the unitary transformation presented in Appendix \ref{appendix:kdphams}, the sixfold node can be brought into block-diagonal form with blocks made from the Hamiltonians describing threefold nodes. For this reason the response of the sixfold node is not fundamentally different from the threefold case considered above.

%
\section{Realistic Material Hamiltonians}\label{sec:realistic_material_hamiltonians}
%

\begin{figure*}  
\begin{center}
\includegraphics[width=0.9\textwidth]{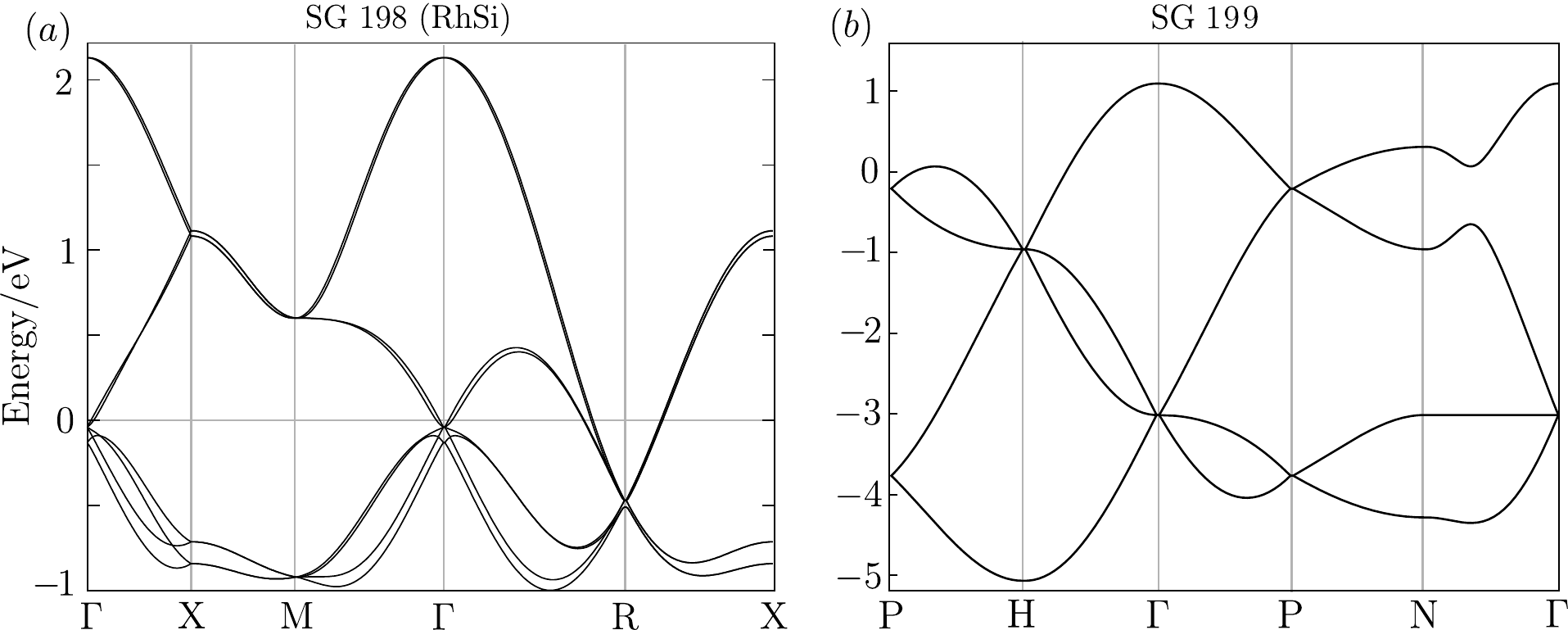}
\caption{\label{fig:TBbands}
{\bf Tight binding band structures}. {\bf (a)} The band structure of the 8-band model used for RhSi, from Ref.~\onlinecite{ChangEA17}, including spin-orbit coupling. The model without spin-orbit coupling is shown in Appendix~\ref{appendix:tight_binding}. {\bf (b)} The band structure of a minimal model featuring the symmetries of space group 199, without spin-orbit coupling. As this is not fit to any particular material the energies are schematic, but the band connectivities and node structures are accurate. 
}
\end{center}
\end{figure*}

To connect our results with realistic materials it is necessary to go beyond low energy effective models, in order to enable us to provide space group specific predictions of the responses we study. 
The tight binding models we present and study in this section incorporate the intrinsic chirality of the space groups and will also take into account the proper embedding of the orbitals in real space. The latter, sometimes overlooked, is strictly necessary to get accurate position operator expectation values and accurate predictions~\cite{Ibanez18}. 

We mainly consider models in two space groups (198 and 199), which can be generalized to three more (212/213 and 214 respectively) by specifying the mentioned orbital real space embedding. The first is space group 198, with minimal $4a$ Wyckoff positions parametrized by a dimensionless number $x$ (see Fig.~\ref{fig:TBbands} (a) for the corresponding band structure, and Appendix~\ref{appendix:tight_binding} for the precise tight-binding model). At $x=1/8$ and $x=5/8$ the symmetry group can be enhanced from tetrahedral to octahedral, and the resulting structure is in the more symmetric groups 212 or 213. Since $x$ can be changed by conjugating the Hamiltonian with a unitary matrix\footnote{although this is not a unitary transformation on the Hilbert space of Bloch functions, since the unitary matrix is not periodic in $\mathbf{k}$.}, the band structure is independent of the embedding. The GME and CPGE responses, however, are sensitive to $x$. The second space group we consider is 199 with minimal $8a$ Wyckoff positions parametrized by $u$. For $u=1/4$, the symmetry can again be promoted from tetrahedral to octahedral, resulting in space group 214. An example band structure for a minimal tight-binding model featuring these symmetries is shown in Fig.~\ref{fig:TBbands}(b). In Appendix~\ref{appendix:tight_binding} we again provide further details of the model band structure.

\begin{figure*}  
\includegraphics[width=\linewidth]{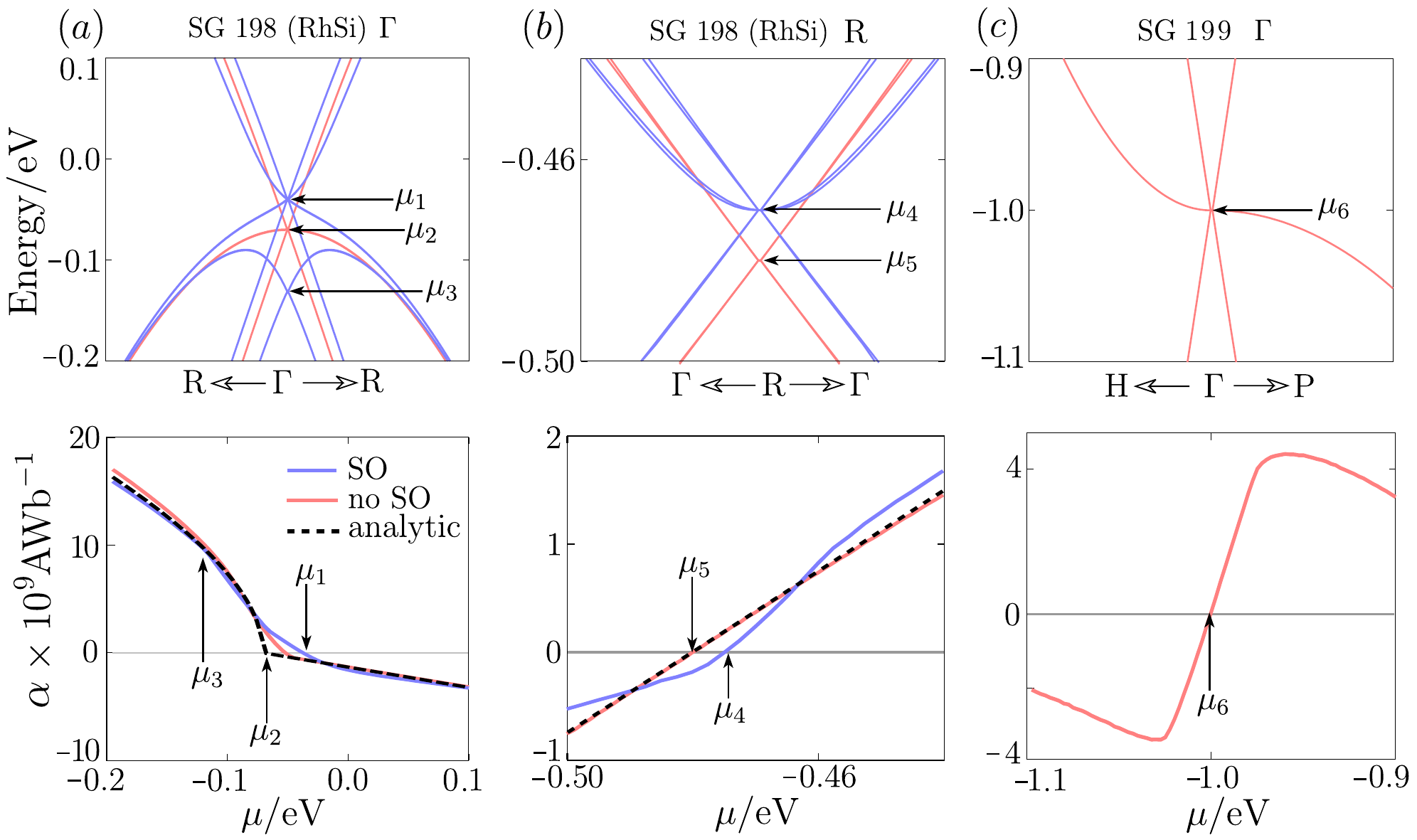}
\caption{\label{fig:RhSi_GME_comparison}
\textbf{Trace of the GME tensor for space groups 198 and 199:} In the top row the different nodes are reproduced from Fig.~\ref{fig:TBbands}. Spin orbit (SO) coupling is either present (blue) or absent (red). In the bottom row the trace of the GME tensor is calculated numerically for the cases with and without spin-orbit coupling (same color scheme), and analytically for the case without spin-orbit coupling (black dashed lines). The chemical potentials of nodes are indicated with arrows. {\bf (a)}  the $\Gamma$ point for RhSi, space group 198. Without spin-orbit coupling the $\Gamma$ point features a doubly-degenerate isotropic threefold node at $-0.07\,$eV ($\mu_2$). The gyrotropic response is linear in chemical potential at energies above the node, and scales as a square root plus a linear part below the node. With spin-orbit coupling the node splits into a fourfold node ($\mu_1$) and a standard Weyl ($\mu_3$). {\bf (b)} the R point in the same model. Without spin-orbit coupling all 8 bands meet at a spin-degenerate double spin-1/2 node ($\mu_5$). Spin orbit coupling splits this into a sixfold node ($\mu_4$) and two separated bands. All cases are isotropic. {\bf (c)} an isolated threefold node appears in space group 199 at the $\Gamma$ point. The node is not isotropic, and the quadratic corrections approximately cancel out leaving a linear variation of the GME trace as a function of chemical potential close to the node.
}
\end{figure*}

Recently, a number of materials in space group 198 have been suggested in which the relevant multifold nodes are predicted to lie near the Fermi level, well separated from other bands. For concreteness we focus on Rhodium Silicide (RhSi), employing the tight-binding band structure developed in Ref.~\onlinecite{ChangEA17}, with modifications specified in Appendix~\ref{appendix:tight_binding}. We plot the band structure in Fig.~\ref{fig:TBbands}. Single crystals of this material have been grown and characterized~\cite{GellerWood54}, with $x=0.3959$ for the relevant bands.

This material features two protected multifold crossings. One of these lies at the $\Gamma$ point. As shown in Fig.~\ref{fig:TBbands} (a) (magnified in Fig.~\ref{fig:RhSi_GME_comparison} (a)), without spin-orbit coupling it takes the form of a spin-degenerate threefold crossing. When spin-orbit coupling is included, the sixfold crossing splits into a fourfold crossing with Chern numbers $C=3,1,-1,-3$, describing a spin-3/2 fermion, and a twofold crossing with a standard Weyl node with $C=1,-1$. At the point R$\,=\left(\pi,\pi,\pi\right)$ without spin-orbit coupling there is a spin-degenerate double spin-1/2 crossing (magnified in Fig.~\ref{fig:RhSi_GME_comparison} (b)). Including spin-orbit coupling splits it into a sixfold crossing and a regular Weyl node~\cite{TangEA17}. In addition, there are several type-II Weyl nodes away from high-symmetry locations~\cite{TangEA17}. The nodes at R and $\Gamma$ have a significant separation in energy. Finally, there is also a double spin-1/2 at M with spin-orbit coupling, however it is far below the Fermi level and plays little role in low-frequency response.

\subsection{GME for Space Groups 198 (RhSi) and 199}\label{sec:GME198tb}

For RhSi we have numerically calculated the GME response, $\alpha_{ij}$, using Eq.~\eqref{eq:alpha} employing the tight-binding model outlined in Ref.~\onlinecite{ChangEA17} with the essential modification described in Appendix~\ref{appendix:tight_binding} to include the real space embedding of the orbitals. This numerical calculation allows us to account for the non-zero spin-orbit coupling, and to move away from the low-energy limit. Calculational details are provided in Appendix~\ref{appendix:GME}.

In Fig.~\ref{fig:RhSi_GME_comparison} we show the contributions to the GME from the $\Gamma$ and R nodes separately, both with and without the spin-orbit coupling present in the real material. The GME tensor $\alpha_{ij}$ is proportional to the identity matrix in all multifold fermions (recall that we denote $\textrm{tr}\left(\alpha_{ij}\right)=\alpha$). We allow the chemical potential to vary in order to give a controllable experimental handle, adjustable by doping.

The $\Gamma$ point without spin-orbit coupling features a spin degenerate threefold node at $-0.07\,$eV. As the chemical potential $\mu$ is lowered above the node, $\alpha$ decreases linearly. At the node the gradient $\textrm{d}\alpha/\textrm{d}\mu$ varies discontinuously, and the trace vanishes at the node (as does the Fermi surface pocket). Below the node there are two contributions to $\alpha$: a linear part, and an $\alpha\propto\sqrt{\mu}$ part. This is in accordance with the analytic predictions of Section \ref{sec:GME_analytics}, overlaid in Fig.~\ref{fig:RhSi_GME_comparison} (See Appendix \ref{sec:AnalyticGME} for details). 

When the spin-orbit coupling is included, the spin degenerate threefold node at $\Gamma$ splits into a fourfold spin-3/2 node and a standard Weyl node, with the separation set by the spin-orbit coupling energy scale. At chemical potentials above the spin-3/2 node, and below the spin-1/2 node, the behavior matches the case without spin-orbit coupling. Aside from affecting the band structure, there is now a spin contribution to the magnetic moment. The effect is around an order of magnitude smaller than the orbital part, and the results presented in Fig.~\ref{fig:RhSi_GME_comparison} are not significantly affected by this term's omission. This is a result of the relatively small spin splitting of the bands around the node in this material.

At the R point without spin-orbit coupling there is a spin degenerate double spin-1/2 node at around $0.48\,$eV below the Fermi level, and the effective Hamiltonian is simply four copies of a Weyl node. Varying $\mu$ about this node leads to a linear change in the GME response, enhanced by a factor of four compared to the standard Weyl case~\cite{ZhongEA16}. The sign of the response switches at the node. 

When spin-orbit coupling is included the spin degenerate double spin-1/2 node at R breaks up into a sixfold (double spin-1) node at $-0.47\,$eV, and two bands separated from this node by the spin-orbit coupling energy scale. The sixfold node is inverted relative to that at $\Gamma$ in the spin-orbit free case.

Also shown in Fig.~\ref{fig:RhSi_GME_comparison} is the response of a threefold node without spin-orbit coupling, as occurs at the $\Gamma$ point in space group 199. This particular realization of a threefold node has anisotropic quadratic corrections about the node, indicated in Fig.~\ref{fig:RhSi_GME_comparison}(c) by showing cuts along the M$\,\,\rightarrow\Gamma\rightarrow\,\,$R directions. The $\Gamma\rightarrow\,\,$X direction features no dispersion close to the node. The quadratic corrections cancel, giving a linear change in the trace of the GME tensor as a function of chemical potential.

As the GME tensor $\alpha_{ij}$ is proportional to the identity in all multifold fermions, the response (either current or magnetization) will always be parallel to the applied (magnetic or electric) field. From the Drude form of the frequency dependence of $\alpha$ (Eq.~\eqref{eq:alpha}), at low frequency in unclean samples the inverse GME is expected to dominate, whereas the direct GME dominates in clean samples at higher frequencies~\cite{ZhongEA16}. Owing to the difficulty with which RhSi and related compounds such as CoSi are grown, it is likely that scattering times $\tau$ will be short. With this in mind, the inverse GME is likely more easily measurable. 

Low and high frequencies are defined relative to the offset of the topological nodes from the Fermi level. In RhSi the node offsets correspond to maximum frequencies of $16.9\,$THz at $\Gamma$ and $116\,$THz at R. The lowest probe frequencies are set by the scattering time $\tau$ in the Drude form of Eq.~\eqref{eq:alpha} and are dependent on sample quality.

It is natural to expect the magnitude of the response from multifold nodes to be significantly enhanced relative to Weyl node pairs, as more bands add to the effect, all contribute with the same sign, and all add either the same amount as a standard Weyl band or some larger multiple thereof. A simple estimate for the GME response in general multifold materials with small spin-orbit coupling can be made by using the coefficients $D_n$ in Table~\ref{table:nodes}. RhSi features one multifold node $0.07\,$eV below $\epsilon_F$ at $\Gamma$, and another $0.48\,$eV below $\epsilon_F$ at R. Owing to the relative sizes of the Fermi surface pockets, the $\Gamma$ node may be neglected. The node at R contributes a (spin-degenerate) double spin-1/2, giving twice the response of a standard (spin-degenerate) Weyl node, featured for example in the candidate chiral Weyl semi-metal SrSi$_2$ in which the node appears around $0.1\,$eV below the Fermi level~\cite{ZhongEA16}. The response in both cases is proportional to the offset of the node from the Fermi level, and this simple estimate therefore suggests a GME response $9.6$ times stronger in RhSi than from a node pair in SrSi$_2$. However, there are multiple symmetry-related nodes in SrSi$_2$ contributing to the material's response. 

For a more thorough estimate we can use the numerically calculated value of $\alpha$ for undoped RhSi in Fig.~\ref{fig:RhSi_GME}, which is around $\alpha=1.3\times 10^{10}\,$AWb$^{-1}$. A $1\,$mT magnetic flux density would therefore generate a current density of around $10^7$Am$^{-2}$. The \emph{rotatory power} of a material is the angle per unit length through which the plane of polarisation of  linearly polarised light is turned upon transmission. It is given in terms of the trace of the GME tensor by~\cite{ZhongEA16}:
\begin{align}
\rho\left(\omega\right)=-\frac{1}{3}\mu_0\alpha\left(\omega\right).
\end{align}
In RhSi at $\omega\rightarrow0$ this gives a value of $5.4\,$rad mm$^{-1}$, an order of magnitude larger than the value of $0.4\,$rad mm$^{-1}$ predicted for a single node pair in SrSi$_2$ for $\omega\rightarrow0$, in agreement with the simple estimate above. 

The GME is not restricted to multifold fermions and Weyl nodes, however, and occurs in any material with a gyrotropic space group and states at the Fermi level. Natural optical activity is also quantified by the rotatory power: the chiral metal tellurium is predicted to exhibit a rotatory power of $0.157(3)\,$rad mm$^{-1}$ at $\hbar\omega=0.117\,$eV and $77\,$K, an order of magnitude smaller than the RhSi response~\cite{SouzaTe}. The optically-active insulator $\alpha$-quartz features a rotatory power of $\rho=0.38\,$rad mm$^{-1}$ at $\lambda=5893\,$\AA, again around an order of magnitude smaller than RhSi~\cite{Nye,ZhongEA93,Arago11}.

\subsection{CPGE for Space Groups 198 (RhSi) and 199}
\begin{figure*}[t]
\begin{center}
\includegraphics[width=\linewidth]{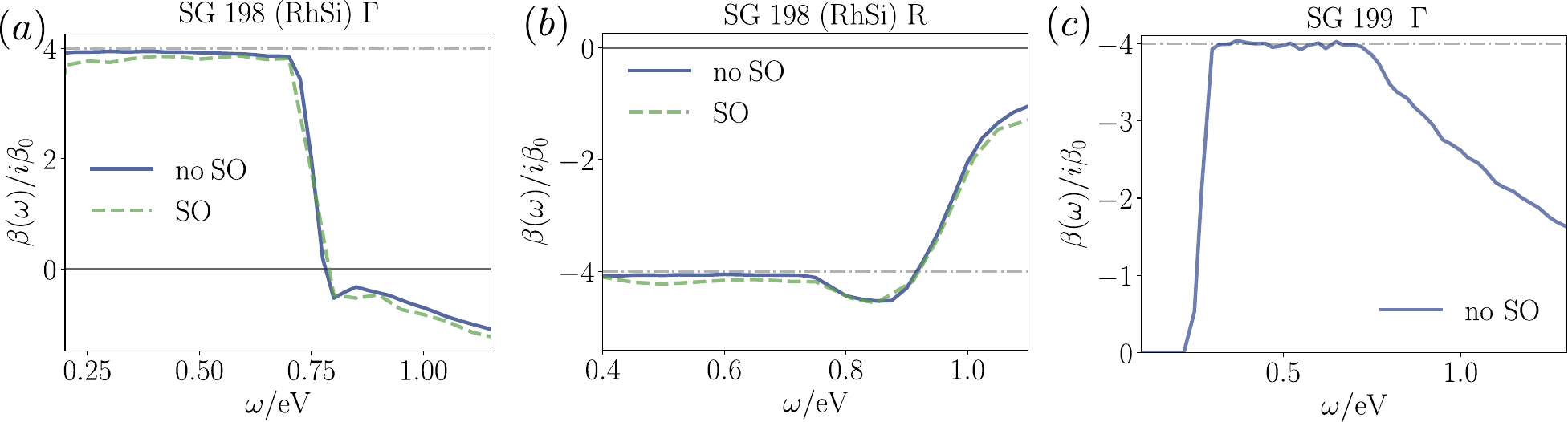}
\caption{\label{fig:SG199_CPGE}
\textbf{Trace of the CPGE tensor for space groups 198 and 199} as a function of frequency $\omega$ for the tight binding models considered in the text. See Fig.~\ref{fig:TBbands} for the corresponding band structures. {\bf (a)} SG198 at $\Gamma$ ($\mu=0.$) {\bf (b)} SG198 at R ($\mu=-0.5$eV) and {\bf (c)} SG199 at $\Gamma$ ($\mu=-2.7$eV). For 198 the CPGE with and without spin-orbit (SO) coupling is plotted. The origin of the observed deviations from quantization  is discussed in the main text and is attributed to a combination of finite momentum space resolution, quadratic corrections and the contributions of small CPGE active pockets. The latter are only allowed in the case with spin-orbit coupling.
}
\end{center}
\end{figure*}

As for the GME we have used the above realistic tight binding lattice models to calculate the CPGE for space groups 198 and 199. The results are summarized in Fig.~\ref{fig:SG199_CPGE}.

There are two main contributions to the CPGE for space group 198, one associated to the $\Gamma$ point and one associated to the R point. Depending on the chemical potential they can either be 
Pauli blocked or optically active, but only in the case where one of them is optically active do we expect the CPGE to be quantized. We analyze both possibilities in what follows.

Illuminating pristine RhSi ($\mu=0$, see Fig.~\ref{fig:TBbands}) will optically activate transitions near the $\Gamma$ point provided the frequency is large enough to overcome Pauli blocking at $\Gamma$ but small enough to Pauli block transitions at R. The corresponding CPGE for frequencies falling in this range is shown in Fig.~\ref{fig:SG199_CPGE} (a). We observe a plateau close to $4\beta_0$ which we interpret as the plateau that corresponds to the optically active threefold (fourfold) fermion at $\Gamma$ without (with) spin-orbit coupling. As explained in section \ref{sec:CPGE_analytics}, the total CPGE is given by the Chern number of the bands with optically allowed transitions, which in both cases (with and without spin-orbit coupling) leads to an expected quantization of $4\beta_0 $, consistent with what we see in the numerical data. It is important to note that the absence of exact quantization is hard to exclude in realistic lattice models since small but finite optical transitions cannot be ruled out. First, there is a Fermi pocket near R which may have allowed optical transitions. Second, the spin-orbit splitting of the degeneracy at $\Gamma$ is small, so changing the frequency slightly can drastically change the nature of the optical surfaces, in particular whether they are open or closed. Finally, quadratic corrections can affect the frequency window at which quantization can occur as discussed in~\ref{sec:CPGE_analytics}.

If instead we assume that the chemical potential can be tuned to be closed to the R point ($\mu=-0.5$ in the energy scale of Fig.~\ref{fig:TBbands})
the relevant transitions will be those around R and the CPGE will resemble that of Fig.~\ref{fig:SG199_CPGE} b) where a plateau close to $-4\beta_0$ appears. In the case without spin-orbit coupling it is the (spin degenerate) fourfold that generates the plateau while in the spin-orbit coupled case it is the sixfold fermion. As for the $\Gamma$ point, the quantization cannot be claimed to be exact, although corrections decrease as the momentum grid size is taken to be finer. 

Our results for space group 198 suggest that the RhSi has indeed, for practical purposes a quantized CPGE at realistic parameter values.

\subsection{GME, CPGE and The Importance of the Orbital Embedding}

\begin{figure}  
\includegraphics[width=\columnwidth]{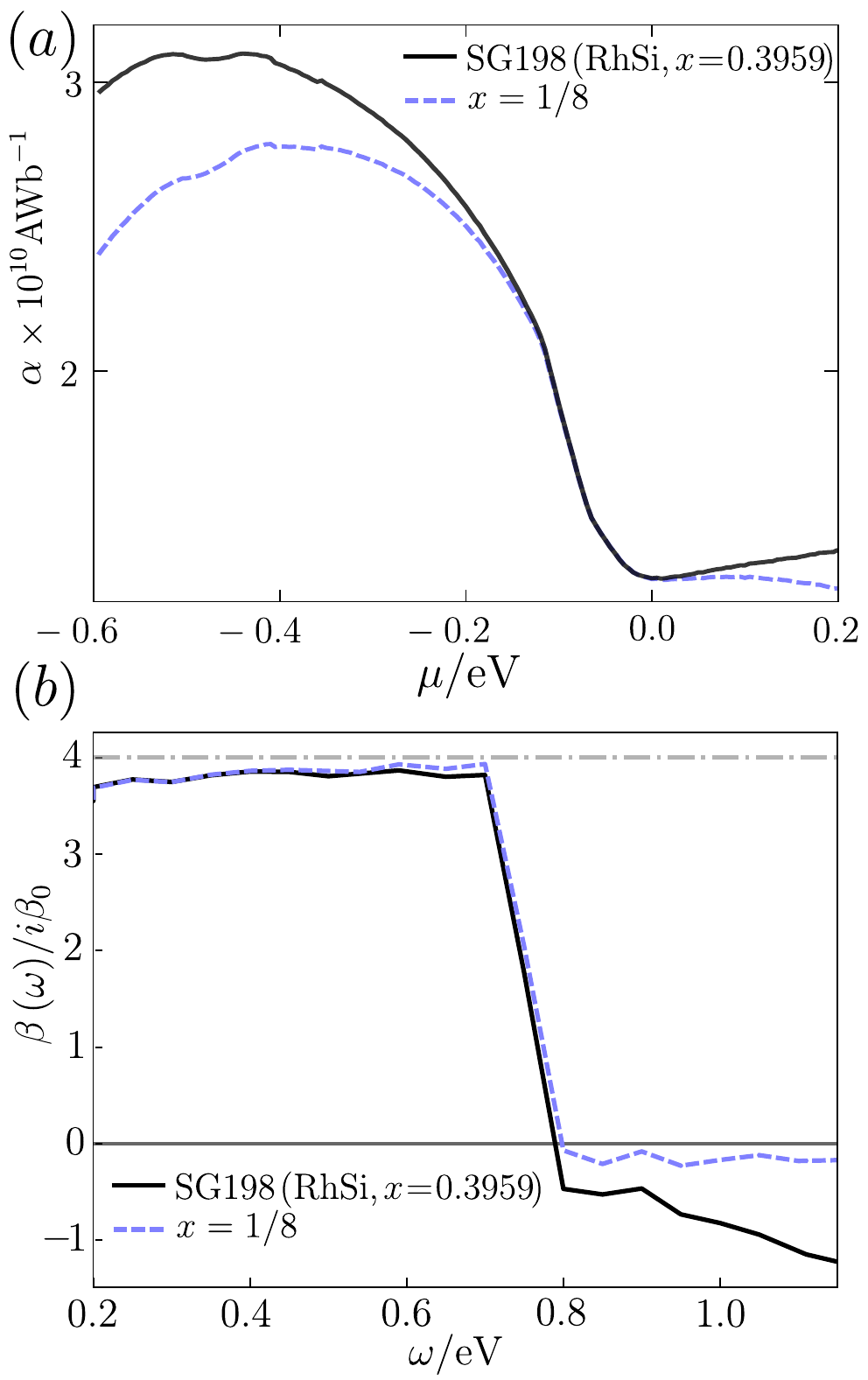} 
\caption{\label{fig:RhSi_GME} 
\textbf{GME and CPGE dependence on atomic positions} {\bf (a)} The trace of the GME tensor in our model of RhSi (black), shown with the trace of the GME tensor in a hypothetical material with $x=1/8$, with the same spectrum but different orbital embeddings (dashed blue). Note the significant deviations in the GME tensor for $\mu$ away from $0$. {\bf (b)} Trace of the CPGE tensor for the same two models. Here we see that the dependence on orbital embedding is strongest for large $\omega$, far from the quantized plateaus.}
\end{figure}

As discussed above, the spatial embedding of the orbitals changes the space group and can change the response functions by modifying the eigenstates.  

In Fig.~\ref{fig:RhSi_GME} we show the total GME and CPGE (for $\mu=0$) responses of the material RhSi as a function of chemical potential and frequency respectively, taking into account spin-orbit coupling. To highlight the effect of the orbital location we also show the response of an imagined material which has an identical band structure, but with $x=1/8$.

For the GME, close to the node around $\epsilon_F$ the two embeddings behave similarly, but away from the node nonlinear corrections become important and the responses differ significantly.  The CPGE also changes for different orbital embeddings, but the effect of $x$ is stronger away from the quantization plateau.

%
\section{Materials predictions}\label{sec:abinitio}
%

\begin{figure*}  
\includegraphics[width=\linewidth]{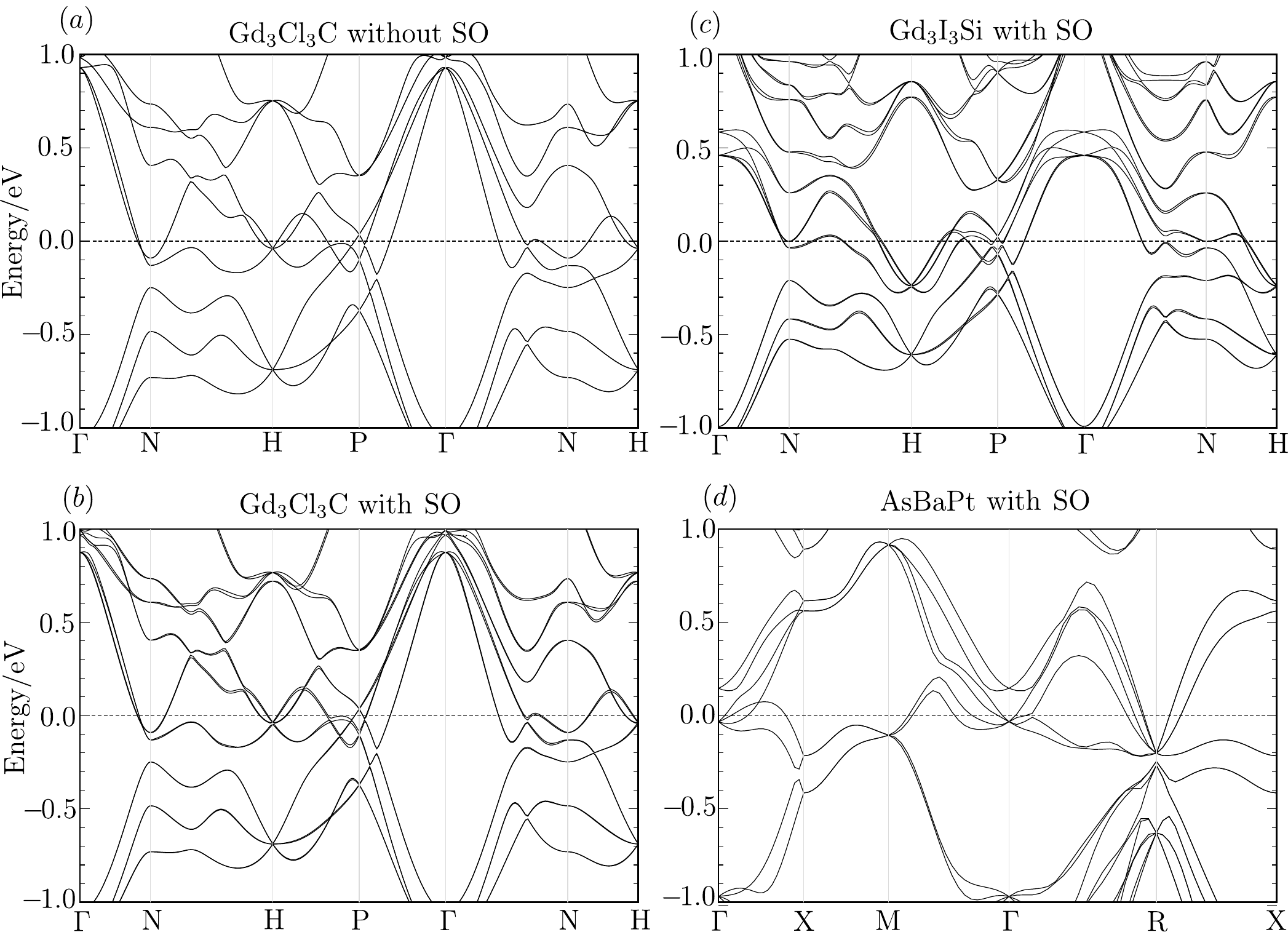}
\caption{\label{fig:abinitio}  
{\bf Band structures of several candidates displaying multifold fermions}.
 {\bf (a)} Gd$_3$Cl$_3$C in SG $I4_{1}32$ (214) without spin-orbit (SO) coupling, featuring a threefold crossing at the H point close to the Fermi level. {\bf (b)} For Gd$_3$Cl$_3$C with spin-orbit coupling included, we observe the same threefold crossing at the H point owing to weak spin-orbit coupling. {\bf c)} Gd$_3$I$_3$Si ($I4_{1}32$) with spin-orbit coupling included.  {\bf (d)} AsBaPt in $P2_{1}3$ (198) with spin-orbit coupling, featuring a fourfold crossing close to the Fermi level at the $\Gamma$ point.
}
\end{figure*}

In order to find material candidates in which the GME and CPGE could be experimentally measured, we have performed an extensive search among the space groups that can display multifold fermions (see Table \ref{table:nodes}). First we searched for spin-degenerate threefold crossings in materials with negligible spin-orbit coupling, and found two material candidates of the same family where threefold crossings are relatively isolated and well split close to the Fermi level. These are Gd$_3$Cl$_3$C\cite{gcc} and Gd$_3$I$_3$Si\cite{gis} in SG $I4_{1}32$ (214) (see Fig.~\ref{fig:abinitio}). In the presence of weak spin-orbit coupling SG 214 displays a threefold crossing at the H point~\cite{manes2012,Cracknell,Elcoro2017}. Figs.~\ref{fig:abinitio} (a) and (b) show the band structure of Gd$_3$Cl$_3$C without and with spin-orbit coupling respectively. We see from the figures that the spin-orbit interaction is indeed weak in this material. At the H point there is still a slight splitting of $4+2$ bands, clearer along high symmetry lines. However, since the splitting is smaller than $1\,$meV, these compounds are suitable for measuring the properties of threefold fermions. 

As was previously reported, sixfold and fourfold fermions are realized in SG $P2_{1}3$ (198) in RhSi. We also present here a new family of ternary compounds in the same SG featuring fourfold fermions near the Fermi level. The principal candidate is AsBaPt\cite{ABP}, shown in  Fig.~\ref{fig:abinitio} (d). One can observe the eightfold connected bands close to the Fermi level, and a fourfold crossing just below the Fermi energy at the $\Gamma$ point. Other candidates of the same family are CaPtSi\cite{CPS}, BaPtSi\cite{BPS}, BaPPt\cite{ABP} and BaPdSi\cite{BPdS}.

%
\section{Conclusions}\label{sec:conclusions}
%

In this work we have calculated two different optical responses of multifold fermions that are enabled by the chiral nature of the space groups they in which they are realized, the gyrotropic magnetic effect (GME) and the circular photogalvanic effect (CPGE). To do so we have first presented a full account and classification of all types of three-fold four-fold and six-fold degeneracies. In particular, we enumerated the space groups and low-energy Hamiltonians for double spin-1/2 degeneracies, which had not been fully specified before.

All multifold nodes can be written in the form $H=\mathbf{k}\cdot\mathbf{S}$, where $\mathbf{k}$ is the crystal momentum, and for certain special parameter choices the matrices $\mathbf{S}$ form representations of the $SU(2)$ algebra. The absence of any characteristic energy scale at low energies and the topological charge of these fermions conspire to produce the peculiar and large responses we find compared to Weyl semimetals and other chiral metals.

Using their low energy description as well as realistic tight binding models for space groups 198 and 199 we have shown that the gyrotropic magnetic effect and the circular photogalvanic effect can serve as experimental probes for distinguishing multifold fermions from other band degeneracies. Multifold (semi-)metals will have an enhanced gyrotropic magnetic effect relative to Weyl semimetals and chiral metals; the GME is the low frequency (transport) limit of natural optical activity, and the rotatory power of multifold materials is similarly enhanced relative to standard cases such as quartz~\cite{Arago11} (although the transparency of quartz means the total optical rotation can be larger in that material). We additionally identified a unique signature of threefold and sixfold nodes in the form of a kink in the GME response as a function of chemical potential about the node.

Furthermore, we have proven under which circumstances multifold fermions can have a quantized CPGE response. This result is surprising, since the corrections spoiling exact quantization of two band models were expected to diverge for bands crossing the nodes~\cite{deJuanEA17}. We have analytically shown that this naive expectation is resolved by virtue of vanishing matrix elements at specific frequency windows where the surfaces of optically allowed transitions are closed. This condition generically results in frequency independent CPGE plateaus. In the specific case where all optical transition surfaces are closed the plateau is strictly quantized in units of $e^3/h^2$ times a sum of Chern numbers. As for the GME, the magnitude of the CPGE is generically larger in multifolds than in Weyl semimetals, since large Chern numbers multiply the large universal constant $\beta_0$ (see Ref.~\onlinecite{deJuanEA17} for a discussion on the estimated size of the photocurrent).

The GME and CPGE therefore can act as multifold detector probes, that distinguish between different types of chiral fermions beyond surface-state ARPES or static magneto-transport, with access, in the case of the CPGE, to their topological charge. It would be interesting to explore how other optical probes, such as resonant X-ray scattering~\cite{K16} which can also carry topological information, can serve as alternatives to the effects discussed in this work.

Throughout this work, we have also had occasion to explore some subtleties related to atomic embeddings which plague the study of both topological semimetals and linear response theory. In our study of space groups 198 and 199, we were careful to emphasize the importance of our basis function positions to our tight binding models. Because the positions of the basis functions do not affect the spectrum of a tight-binding Hamiltonian, they are often chosen arbitrarily or overlooked entirely. We have emphasized in our discussion that the basis function positions, due to their effect on the boundary conditions of Bloch functions, have measurable consequences. In particular, both the GME and CPGE depend on matrix elements of the position operator, which depend crucially on these choices~\cite{Ibanez18}. In space groups 198 and 199, the basis function positions are not constrained by symmetry, and we showed that different choices can have a marked effect on the non-quantized part of response functions. We hope this pedagogical exercise serves as a useful guide to future research on response theory.

While both the GME and CPGE provide implementable tests for the existence and properties of multifold nodes, there are some drawbacks and experimental challenges to be overcome. First, both effects require chiral, non-magnetic, metallic multifold systems in order to be measured. Such materials are quite rare yet this adds value to the family of materials we have presented in this work. The potential of recently developed efficient approaches\cite{TQC,Po2017,Song2017} for the search of novel topological metals makes us confident that other materials can be soon added to the list where these effects can be measured.

Second, although the GME has been measured~\cite{Fukuda75,Vorobev79,Shalygin2012,Furukawa2017} and largely understood~\cite{SouzaTe,Sahin18} in materials like Tellurium, it so far lacks experimental confirmation in topological metals. We believe our work brings it closer to experimental realization, since we have showed that multifold fermions add additional materials where the GME can be probed and measured due its large magnitude and the multiple materials that can potentially display the effect.

Thirdly, measurements of the bulk injection current in the CPGE may require some experimental ingenuity. As discussed in \onlinecite{deJuanEA17} we expect that quantization can have its clearest signatures in a time-resolved photocurrent measurement, with the use of light pulses that are shorter than the relevant scattering time $\tau\sim $ps. Measuring the photo-current directly in topological semimetals is possible~\cite{Ma17,Sun17,Burch17,Rappe18}, and although it will result in CPGE plateaus, these will generically depend on the scattering $\tau$ and not only on fundamental constants \cite{deJuanEA17,Konig17}. 

Finally, as we have seen the quantization of the CPGE can be weakly violated at certain frequencies due to the presence of quadratic corrections and extra optical transitions, necessitating some care in choosing the appropriate experimental platform. Additionally, measuring the trace of the CPGE tensor $\beta_{ij}$ requires summing over all polarization planes which might be experimentally challenging in chiral Weyl semimetals. Remarkably, all multifolds except doubled spin-1/2 occur in cubic point groups, so $\beta_{ij} = \delta_{ij}$ and measuring one component is enough.
This presents a clear practical advantage over non-cubic Weyl semimetals~\cite{deJuanEA17}.

Weyl semi-metals provide a condensed matter analogue of exotic (and as yet unobserved) fundamental particles, Weyl fermions, governed by the Weyl equation of particle physics. Multifold fermions provide a condensed matter analogue of particles `beyond the standard model'. In this work we have provided their full classification and unraveled, both analytically and numerically their gyrotropic and photogalvanic responses, providing as well new materials where these predictions can be tested.  Our results aim to motivate experimental work and material growth that may lead to a deeper understanding of these emergent pseudo-relativistic excitations as well as enhanced optical phenomena. 

\section{Acknowledgments}

The authors wish to thank K. Burch for providing helpful comments on the manuscript. BB acknowledges helpful discussions with Ivo Souza and Benjamin Wieder, and the support of the Donostia International Physics Center, where a portion of this work was completed. BB and AGG additionally acknowledge the hospitality of the Banff International Research Station. FF acknowledges funding from the Astor Junior Research Fellowship of New College, Oxford. FdJ and AGG acknowledge funding from the European Union's Horizon 2020 research and innovation programme under the Marie-Sklodowska Curie grant agreements No. 705968 (FdJ) and 653846 (AGG). TM was supported by the Gordon and Betty Moore Foundation’s EPiQS Initiative Theory Center Grant to UC Berkeley, and the Quantum Materials program at LBNL, funded by the US Department of Energy under Contract No. DE-AC02-05CH11231. MGV was supported by IS2016-75862-P national project of the Spanish MINECO. AGG acknowledges M.A. Sanchez-Martinez for discussions and collaboration on related work. First principles calculations were performed on the Atlas supercomputer of the Donostia International Physics Center.

\bibliographystyle{apsrev4-1}
\bibliography{mybib}

%
\newpage
\pagebreak
\appendix

\counterwithin{figure}{section}

%
\begin{widetext}
\section{Calculational Details to derive the GME}
\label{appendix:GME}
%

%
\subsection{Mathematical Details of Gyrotropy and Connection to GME}

In general materials the displacement field $D_i$ and electric field $E_j$ can be related through the expression~\cite{AG84,Nye}
\begin{equation}
\label{eq:D}
	D_i = \left(\varepsilon_{ij}(\omega) -i \lambda_{ijl}(\omega)k_l+\cdots\right)E_j,
\end{equation}
where $\epsilon_{ij}$ is the dielectric susceptibility, and $\lambda_{ijl}\left(\omega\right)$ is the gyrotropy tensor. The material is optically active (gyrotropic) when the gyrotropy tensor is nonzero. The ellipsis accounts for the fact that this relation can be viewed as an expansion in the momentum of the incident radiation, which is typically small compared to the inverse lattice spacing. For notational convenience, we will suppress the dependence of tensors on angular frequency $\omega$ for the remainder of the discussion. A non-zero $\lambda$ implies that inversion symmetry is broken. Moreover since $\varepsilon_{ij}(\omega,\mathbf{k})=\varepsilon_{ji}(\omega,-\mathbf{k})$ k we have that $\lambda_{ijl}=-\lambda_{jil}$. Therefore $\lambda_{ijl}$ has 9 independent components and can be written as $\lambda_{ijl}=\epsilon_{ijm}g_{ml}$ in terms of the gyrotropy tensor $g_{ml}$, with $\epsilon_{ijm}$ the Levi-Civita symbol. This enables us to define the gyration vector $\mathbf{G}$ through 
\begin{equation}
\lambda_{ijl}k_l= \epsilon_{ijm}g_{ml}k_l \equiv \epsilon_{ijm}G_m,
\end{equation}
and thus Eq.~\eqref{eq:D} can be written as
\begin{equation}\label{eq:D_G}
	D_i = \varepsilon_{ij}E_j -i (\mathbf{G}\times \mathbf{E})_i.
\end{equation}
Note that the GME tensor $\alpha_{ij}$ used in the main text is related to the gyrotropy tensor through the relation
\begin{equation}
g_{ij}=\frac{1}{\omega c\epsilon_0}\left(\alpha_{ji}-\textrm{tr}\left(\alpha\right)\delta_{ij}\right),
\end{equation}
as stated in the supplementary material of Ref.~\onlinecite{ZhongEA16}.
It is sometimes custom to invert Eq.~\eqref{eq:D_G} and write
\begin{equation}
	E_i = \varepsilon^{-1}_{ij}D_j -i (\mathbf{f}\times \mathbf{D})_i,
\end{equation}
where we have implicitly defined
\begin{equation}
	\varepsilon^{-1}_{ij}(\omega,\mathbf{k})=\varepsilon^{-1}_{ij}(\omega)+i\delta_{ijl}k_l\cdots,
\end{equation}
and
\begin{equation}
	\delta_{ijl}k_l = \epsilon_{ijm}f_{ml}k_l \equiv \epsilon_{ijm}f_m.
\end{equation}\\
Note that $\mathbf{f}$ has the same symmetry properties as $\mathbf{G}$. 
The inverse gyroptropy tensor $f_{ij}$ only enters the electromagnetic wave equations with the scalar product $f_{ij}\hat{k}_i\hat{k}_j$~\cite{AG84}. The rotatory power, defined as the angle of rotation of the plane of polarization per unit length of propagation, is proportional to $f_{ij}\hat{k}_i\hat{k}_j$ and is thus only determined by the symmetric part of $f_{ij}$.
For the point groups $C_{4v,6v,3v}$ the second rank tensor $g_{ij}$ and thus $f_{ij}$ have a zero symmetric part, and thus do not rotate the plane of polarization of light. They 
can however be `weakly gyrotropic', meaning that the gyrotropy tensor is nonzero but purely antisymmetric.
The symmetry properties of the inverse gyrotropy tensor are shown in Table~\ref{table:symmetry}.

The tensor $\alpha_{ij}$, used in the main text, is given by
\begin{equation}
j_{i} = \alpha_{ij}B_{j}
,\end{equation}
which, together with Eq.~\eqref{eq:D}, implies that $\alpha_{ij}$ shares the same symmetry properties as $g_{ij}$.
This extra $i/\omega$ prefactor differs from the definitions used in Ref.~\onlinecite{MaPesin15}.

\subsection{Numerical Evaluation of the GME Tensor}

Reference~\onlinecite{ZhongEA16} gives the following expression for the GME tensor $\alpha_{ij}$
\begin{align}
\alpha_{ij}=\frac{e}{\left(2\pi\right)^{3}}\sum_{n}\int\text{d}^{3}\boldsymbol{k}\frac{\partial f}{\partial\epsilon_{n}}v_{ni}\left(\frac{e}{2\hbar}\mathfrak{Im}\epsilon_{jlm}\sum_{n'\ne n}\frac{\langle n|\partial_{l}H_{\boldsymbol{k}}|n'\rangle\langle n'|\partial_{m}H_{\boldsymbol{k}}|n\rangle}{\epsilon_{n}-\epsilon_{n'}}-\frac{eg_{s}\hbar}{4m_{e}}\langle n|\sigma_{j}|n\rangle\right).
\end{align}
Here, $e$ is the electron charge; $f\left(\epsilon_{n}\right)$ is the Fermi function evaluated for band $n$ with energy $\epsilon_{n}$;
$v_{ni}=\hbar^{-1}\partial_{i}\epsilon_{n}=\langle n|\hbar^{-1}\left(\partial_{i}H_{\boldsymbol{k}}\right)|n\rangle$
is the velocity of band $n$ in Cartesian direction $i$, where $\partial_{i}\triangleq\partial/\partial k^{i}$; $g_{s}\approx2$ is the spin g-factor for the electron; $m_{e}$ is the mass of the electron; and $\sigma_{j}$ is the $j^{\textrm{th}}$ Pauli matrix. Einstein summation notation is assumed.

Evaluating the expression at zero temperature, the Fermi function
reduces to a Dirac delta function
\begin{align}
\alpha_{ij}=-\frac{e}{\left(2\pi\right)^{3}}\sum_{n}\int\text{d}^{3}\boldsymbol{k}\delta\left(\epsilon_{n}-\mu'\right)v_{ni}\left(\frac{e}{2\hbar}\mathfrak{Im}\epsilon_{jlm}\sum_{n'\ne n}\frac{\langle n|\partial_{l}H_{\boldsymbol{k}}|n'\rangle\langle n'|\partial_{m}H_{\boldsymbol{k}}|n\rangle}{\epsilon_{n}-\epsilon_{n'}}-\frac{eg_{s}\hbar}{4m_{e}}\langle n|\sigma_{j}|n\rangle\right).
\end{align}

We now employ the following approximation to deal with the delta function
\begin{align}
\int\text{d}^{3}\boldsymbol{k}\delta\left(\epsilon_{n}-\epsilon_{F}\right) & \approx\frac{1}{\delta E}\int_{\epsilon_{F}-\delta E/2}^{\epsilon_{F}+\delta E/2}\text{d}E'\int\text{d}^{3}\boldsymbol{k}H\left(E'\right),
\end{align}
where
\begin{align}
H\left(E\right)= & \begin{cases}
1\,\,\text{if} & \epsilon_{F}-\frac{\delta E}{2},<\epsilon_{n}<\epsilon_{F}+\frac{\delta E}{2}\\
0\,\, & \text{otherwise}.
\end{cases}
\end{align}
The $\delta E$ expression must be symmetric about $\epsilon_{F}$. This
can be seen from the GME expression for a single node, which is proportional
to $\left(\epsilon_{node}-\epsilon_{F}\right)$; if the window
of $H\left(E\right)$ were defined to be $\epsilon_{F}<\epsilon<\epsilon_{F}+\delta E$, the average value of $\epsilon_{node}$ would become proportional to $\delta E$,
leading to a linear dependence of $\alpha$ on the choice $\delta E$. Physically, $\delta E$ corresponds to a finite-width shell of energies to be averaged over around the Fermi level.

Defining the volume of the crystal $V=N^{3}a^{3}$ we have, in partial pseudocode
\begin{align}
\alpha_{ij}=-\frac{e}{V}\frac{1}{\delta E}\sum_{n}\sum_{\boldsymbol{k}\in BZ}\left[\text{if }-\frac{\delta E}{2}<\epsilon_{\boldsymbol{k}n}<\frac{\delta E}{2}\right]\left[\frac{1}{\hbar}\langle n|\partial_{i}H_{\boldsymbol{k}}|n\rangle\right]\left(\frac{e}{2\hbar}\mathfrak{Im}\epsilon_{jlm}\sum_{n'\ne n}\frac{\langle n|\partial_{l}H_{\boldsymbol{k}}|n'\rangle\langle n'|\partial_{m}H_{\boldsymbol{k}}|n\rangle}{\epsilon_{n}-\epsilon_{n'}}-\frac{eg_{s}\hbar}{4m_{e}}\langle n|\sigma_{j}|n\rangle\right).
\end{align}
Now define the dimensionless variable $k=\frac{2\pi}{a}\overline{k}$,
with $a$ the lattice constant of the material, such that the Brillouin
zone is defined by $\overline{k}\in\left[-\frac{1}{2},\frac{1}{2}\right]$. Using $g_s=2$:
\begin{align}
\alpha_{ij}=-\frac{e^{2}}{4\pi h^{2}}\frac{1}{\delta E}\sum_{n}\left[\frac{1}{N^{3}}\sum_{\boldsymbol{\overline{k}}\in BZ}\right]\left[\text{if }-\frac{\delta E}{2}<\epsilon_{\boldsymbol{k}n}<\frac{\delta E}{2}\right]\langle n|\overline{\partial}_{i}H_{\boldsymbol{\overline{k}}}|n\rangle\left(\mathfrak{Im}\epsilon_{jlm}\sum_{n'\ne n}\frac{\langle n|\overline{\partial}_{l}H_{\overline{\boldsymbol{k}}}|n'\rangle\langle n'|\overline{\partial}_{m}H_{\boldsymbol{\overline{k}}}|n\rangle}{\epsilon_{n}-\epsilon_{n'}}-\frac{h^{2}}{m_{e}a^{2}}\langle n|\sigma_{j}|n\rangle\right).
\end{align}

Note that the only material-specific parameter is the lattice
constant, which only affects the spin part. In the main text, the only
material whose spin-split bandstructure we consider is RhSi, in Section~\ref{sec:realistic_material_hamiltonians}.
Inserting the values of fundamental constants and the RhSi lattice
constant $a=4.67\,\textrm{\AA}$ for the spin part, we have the final result (with $\overline{E}$ \emph{etc.} dimensionless):
\begin{align}
\alpha^{\mathrm{RhSi}}_{ij}	=-7.45\times10^{8}\frac{\text{A}}{\text{Tm}^{2}}\frac{1}{\overline{\delta E}}\sum_{n}\left[\frac{1}{N^{3}}\sum_{\boldsymbol{\overline{k}}\in BZ}\right]\left[\text{if }\!\!-\!\frac{\overline{\delta E}}{2}\!<\!\epsilon_{\boldsymbol{k}n}\!<\!\frac{\overline{\delta E}}{2}\right]\langle n|\overline{\partial}_{i}\overline{H}_{\boldsymbol{\overline{k}}}|n\rangle\Biggl(\left[\frac{1}{\text{eV}}\right]\mathfrak{Im}\epsilon_{jlm}\sum_{n'\ne n}\frac{\langle n|\overline{\partial}_{l}\overline{H}_{\overline{\boldsymbol{k}}}|n'\rangle\langle n'|\overline{\partial}_{m}\overline{H}_{\boldsymbol{\overline{k}}}|n\rangle}{\overline{\epsilon}_{n}-\overline{\epsilon}_{n'}}\nonumber\\
	-13.8\langle n|\sigma_{j}|n\rangle\Biggr).
\end{align}

\subsection{Analytic Evaluation of the Orbital Magnetic Moment}\label{sec:AnalyticGME}

We illustrate the derivation of the orbital magnetic moment, required for the GME calculations, for the case of the threefold node. This also applies to the sixfold case by the reasoning presented in the main text. The threefold node, as described by Eq.~\eqref{eq:H3f}, with $\phi=\pi/2$ (general expressions for $\phi\neq \pi/2$ can be found in Appendix~\eqref{appendix:H3f}), reads
\begin{align}
H=i\hbar v_{F}\left(\begin{array}{ccc}
0 & k_{x} & -k_{y}\\
-k_{x} & 0 & k_{z}\\
k_{y} & -k_{z} & 0
\end{array}\right)
\end{align}
with eigenenergies
\begin{align}
E_{1}=0,\,\,E_{2}=\hbar v_{F}k,\,\,E_{3}=-\hbar v_{F}k
\end{align}
and corresponding normalized eigenvectors
\begin{eqnarray}
|1\rangle &=& \dfrac{1}{k}\left(k_{z},k_{y},k_{x}\right)^{T}\nonumber\\
|2\rangle &=&\left(\sqrt{2}k\sqrt{k_{x}^{2}+k_{z}^{2}}\right)^{-1}\left(\begin{array}{c}
k_{y}k_{z}- ikk_{x}\nonumber\\
-k_{x}^{2}-k_{z}^{2}\\
k_{x}k_{y}+ ikk_{z}
\end{array}\right)\\
|3\rangle &=&\left(\sqrt{2}k\sqrt{k_{x}^{2}+k_{z}^{2}}\right)^{-1}\left(\begin{array}{c}
k_{y}k_{z}+ ikk_{x}\\
-k_{x}^{2}-k_{z}^{2}\\
k_{x}k_{y}- ikk_{z}
\end{array}\right).
\end{eqnarray}

The Berry curvature is given by
\begin{align}
\Omega_{i}^{n}=i\epsilon_{ijk}\sum_{m\ne n}\frac{\langle n|\left(\partial_{j}H\right)|m\rangle\langle m|\left(\partial_{k}H\right)|n\rangle}{\left(E_{n}-E_{m}\right)^{p}}
\end{align}
with $p=2$. With $p=1$ the expression instead gives the orbital magnetic moment divided by $e/2\hbar$. 

\noindent For band $n=1$ we have:

\begin{eqnarray}
\Omega_{1}^{1}&=&i\frac{\langle1|\partial_{2}H|2\rangle\langle2|\partial_{3}H|1\rangle}{\left(E_{1}-E_{2}\right)^{2}}+i\frac{\langle1|\partial_{2}H|3\rangle\langle3|\partial_{3}H|1\rangle}{\left(E_{1}-E_{3}\right)^{2}}+c.c.\nonumber\\
&=&i\frac{\langle1|\partial_{2}H|2\rangle\langle2|\partial_{3}H|1\rangle}{\left(-\hbar v_{F}k\right)^{2}}+i\frac{\langle1|\partial_{2}H|3\rangle\langle3|\partial_{3}H|1\rangle}{\left(\hbar v_{F}k\right)^{2}}+c.c.\nonumber\\
&=&\left(\hbar v_{F}\right)^{2}\frac{\left(\sqrt{2}k\right)^{-1}\left(k_{z}^{2}+k_{x}^{2}\right)^{1/2}\left(\sqrt{2}k\sqrt{k_{x}^{2}+k_{z}^{2}}\right)^{-1}\left(k_{x}k-ik_{z}k_{y}\right)}{\left(-\hbar v_{F}k\right)^{2}}+\nonumber\\
&+&\left(\hbar v_{F}\right)^{2}\frac{\left(\sqrt{2}k\right)^{-1}\left(k_{z}^{2}+k_{x}^{2}\right)^{1/2}\left(\sqrt{2}k\sqrt{k_{x}^{2}+k_{z}^{2}}\right)^{-1}\left(-k_{x}k-ik_{z}k_{y}\right)}{\left(\hbar v_{F}k\right)^{2}}+c.c.
\end{eqnarray}
The real parts of the first two terms cancel, leaving the imaginary parts, which then cancel with the complex conjugates, giving zero
\begin{align*}
\Omega_{i}^{1}=0
\end{align*}
(the other components similarly vanish owing to the isotropic nature of the Weyl node). 

To obtain the orbital magnetic moments we set $p=1$ rather than $2$. As a result, a relative sign is introduced between the formerly canceling terms. The result is therefore
\begin{eqnarray}
m_{1}^{1}&=&\frac{e}{2\hbar}\left(\hbar v_{F}\right)^{2}\frac{\left(\sqrt{2}k\right)^{-1}\left(k_{z}^{2}+k_{x}^{2}\right)^{1/2}\left(\sqrt{2}k\sqrt{k_{x}^{2}+k_{z}^{2}}\right)^{-1}\left(k_{x}k-ik_{z}k_{y}\right)}{-\hbar v_{F}k}+\nonumber\\
 &+&\frac{e}{2\hbar}\left(\hbar v_{F}\right)^{2}\frac{\left(\sqrt{2}k\right)^{-1}\left(k_{z}^{2}+k_{x}^{2}\right)^{1/2}\left(\sqrt{2}k\sqrt{k_{x}^{2}+k_{z}^{2}}\right)^{-1}\left(-k_{x}k-ik_{z}k_{y}\right)}{\hbar v_{F}k}+c.c.\nonumber\\
&=&-ev_{F}k^{-2}k_{x}
\end{eqnarray}
and similarly
\begin{align}
m_{i}^{1}=-ev_{F}k^{-2}k_{i}.
\end{align}

\noindent Bands $n=2,\,3$:

\begin{align}
\Omega_{1}^{2}=i\frac{\langle2|\left(\partial_{2}H\right)|1\rangle\langle1|\left(\partial_{3}H\right)|2\rangle}{\left(E_{2}-E_{1}\right)^{2}}+i\frac{\langle2|\left(\partial_{2}H\right)|3\rangle\langle3|\left(\partial_{3}H\right)|2\rangle}{\left(E_{2}-E_{3}\right)^{2}}+c.c.
\end{align}
the second term vanishes, since $\langle2|=|3\rangle^{T}$ and the $\partial_{i}H$ are elements of the cross product.

\begin{eqnarray}
\Omega_{1}^{2}&=&-\left(\hbar v_{F}\right)^{2}\frac{\left(\sqrt{2}k\right)^{-2}\left(ik_{z}k_{y}+k_{x}k\right)}{\left(\hbar v_{F}k\right)^{2}}+c.c.\nonumber\\
&=&-\left(\hbar v_{F}\right)^{2}\frac{k^{-1}k_{x}}{\left(\hbar v_{F}k\right)^{2}}
\end{eqnarray}
giving
\begin{align}
\Omega_{i}^{2,3}=\left(-,+\right)k^{-3}k_{i}.
\end{align}

Again changing the denominator in the penultimate line we obtain the expression for the orbital magnetic moment:
\begin{align}
m_{i}^{2,3}=-ev_{F}k^{-2}k_{i}
\end{align}
(N.B. this has the same sign between bands). The other cases can be calculated in a similar fashion.

With the calculated orbital moments, the contribution to the GME trace for band $n$ can be computed as 
\begin{align}
\alpha_n &=  e \int \frac{d^3k}{(2\pi)^3} \frac{\partial_{k_i}E_n}{\hbar} m_{i,n} \delta(E_n - \mu) \nonumber\\
&= \frac{e}{\hbar} \int \frac{d^3k}{(2\pi)^3} \partial_{k_i}E_n m_{i,n} \frac{\delta(k-k_F)}{|\partial_{k_i}E_n|} \nonumber \\
&= \frac{e}{ 2 \pi^2 \hbar}  \int k^2dk  \frac{e v_F D_n}{2 k} \delta(k-k_F)\nonumber \\
&= \frac{e^2}{ 4 \pi^2 \hbar} k_F  v_F D_n 
\end{align}

For the threefold fermion at the $\phi=\pi/2$ point at chemical potential $\mu$ measured from the node, the upper band has $k_F = \mu/(\hbar v_F)$. For the middle band, quadratic corrections are needed to have a Fermi surface. Including $H = 1/2m (k_x^2+k_y^2+k_z^2)$ which does not change the orbital moment, we have $k_F = \hbar^{-1}\sqrt{2m \mu}$. So for the three bands we have
\begin{align}
\alpha_n = \frac{e^2}{ 4 \pi^2 \hbar^2} \left\{ \begin{array}{c}  \mu D_1 \Theta(\mu) \\ v_F \sqrt{2m \mu} D_2 \Theta(\mu) \\ |\mu|D_3 \Theta(-\mu).\end{array} \right.
\end{align}

To compare this prediction with the tight binding model for RhSi used in Section \ref{sec:GME198tb}, we need to compute $v_F$ and $m$ from the tight binding parameters $v_1$, $v_p$, and $v_2$ of Ref.~\onlinecite{ChangEA17}. These can be obtained in perturbation theory to be $v_F = v_p/2$ and $m=\tfrac{1}{2} [(4 v_1^2 - v_p^2)/(16 v_1) - v_2/2]^{-1}$.

\begin{table*}[hbt]
  \begin{tabular}{l|c|c|c|c|c|c|c|c|c|c}
    Group & P & E  & G & wG & OA & Pz & LPGE & CPGE & $f_{ij}^s$ \\
    \hline
    $C_1$ & $\surd$ & $\surd$ & $\surd$ &  & $\surd$ & $\surd$ & $\surd$ & $\surd$ & all \\
    $C_2$ & $\surd$ & $\surd$ & $\surd$ & &$\surd$ & $\surd$ & $\surd$& $\surd$ & $f_{11},\,f_{22},\, f_{33},\,f_{13}$ \\
    $C_s$ & $\surd$&  & $\surd$& & $\surd$& $\surd$& $\surd$& $\surd$ & $f_{12},\,f_{23}$ \\
    $C_{2v}$ & $\surd$& & $\surd$& &$\surd$ & $\surd$& $\surd$& $\surd$ & $f_{12}$ \\
    $C_{3}$ & $\surd$& $\surd$ & $\surd$& &$\surd$ & $\surd$& $\surd$& $\surd$ & $f_{11}=f_{22},\, f_{33}$ \\
    $C_{3v}$ & $\surd$& &$\surd$ &$\surd$ & & $\surd$& $\surd$& $\surd$ & none \\
    $C_4$ &$\surd$ & $\surd$ & $\surd$ & & $\surd$& $\surd$& $\surd$& $\surd$ & $f_{11}=f_{22},\, f_{33}$ \\
    $C_{4v}$ &$\surd$ & & $\surd$ & $\surd$ & & $\surd$ & $\surd$& $\surd$ & none \\  
    $C_6$ &$\surd$ & $\surd$ & $\surd$& &$\surd$ & $\surd$& $\surd$& $\surd$ & $f_{11}=f_{22},\, f_{33}$ \\  
    $C_{6v}$ & $\surd$& &$\surd$ & $\surd$ & & $\surd$& $\surd$& $\surd$ & none \\
    $D_2$ & & $\surd$&$\surd$ & &$\surd$ & $\surd$& $\surd$& $\surd$ & $f_{11},\,f_{22},\,f_{33}$ \\
    $D_{2d}$ & & &$\surd$ & & $\surd$& $\surd$& $\surd$& $\surd$ & $f_{12}$ \\
    $D_3$ & & $\surd$& $\surd$& & $\surd$&$\surd$ & $\surd$& $\surd$ & $f_{11}=f_{22},\,f_{33}$ \\  
    $D_4$ & & $\surd$& $\surd$& & $\surd$& $\surd$& $\surd$& $\surd$ & $f_{11}=f_{22},\,f_{33}$ \\    
    $D_6$ & & $\surd$&$\surd$ & & $\surd$&$\surd$& $\surd$& $\surd$ & $f_{11}=f_{22},\,f_{33}$ \\
    $S_4$ & & & $\surd$& & $\surd$& $\surd$& $\surd$ & $\surd$ & $f_{11}=-f_{22},\,f_{12}$ \\
    $C_{3h}$ & & & & & & $\surd$& $\surd$ & & none \\
    $D_{3h}$ & & & & & & $\surd$& $\surd$& & none \\
    $T_d$ & & & & & & $\surd$& $\surd$& & none \\  
    $T$ & & $\surd$& $\surd$ & & $\surd$&$\surd$& $\surd$ &$\surd$ & $f_{11}=f_{22}=f_{33}$ \\
    $O$ & & $\surd$& $\surd$& & $\surd$& &  & $\surd$ & $f_{11}=f_{22}=f_{33}$ \\
  \end{tabular}
  \caption{\label{table:symmetry} 
  List of non-centrosymmetric point groups, \emph{i.e.} groups without an inversion center. The tick indicates if the point group is Polar (P), Enantiomporphic (E), Gyrotropic (G), weakly Gyrotropic (wG), Piezoelectric (Pz), Optically active (OA) or has a linear or circular photogalvanic effect (LPGE and CPGE). The trace of the CPGE or GME is only nonzero for enantiomorphic (E) groups. The final column lists the nonzero elements of the symmetric part of the inverse gyrotropy tensor $f_{ij}$, $f_{ij}^s=f_{ji}^s$, which has the same symmetry as $g_{ij}$ and $\alpha_{ij}$. Parts of this table appear in Refs.~\onlinecite{Fedorov59,Fedorov73,Eritsyan82,AG62,AG84,Nye}.
  }
\end{table*}

%
%
\section{Symmetry Constraints for Response Coefficients in non-Centrosymmetric Point Groups}
\label{appendix:symmetries}
%

\emph{Non-centrosymmetric point groups} -- There exist 21 noncentrosymmetric point groups, listed in Table \ref{table:symmetry}. In the following we discuss some of the effects they can host due to the absence of inversion and list the corresponding subset of point groups for each effect.

\emph{Polar point groups} -- There are 10 polar point groups (also known as ferro-electric or pyro-electric materials): 
$$C_1,C_s,C_2,C_{2v},C_4,C_{4v},C_3,C_{3v},C_6,C_{6v}.$$
In a polar point group all symmetries, including mirrors, leave one direction invariant. An insulator with these symmetries can have a non-vanishing polarization along this direction. 

\emph{Enantiomorphic or chiral groups} -- These groups, defined as those with no orientation-reversing elements, are~\cite{Newnham}: 
$$C_1,C_2,C_3,C_4,C_6,D_2,D_4,D_6,T,O.$$
In this work, we have considered the following space groups with chiral point groups that can host different multifold fermions: 18, 19 ($D_2$), 90, 92, 94, 96 ($D_4$), 195-199 ($T$), 207-214 ($O$). The constraints for the gyrotropy tensor (and thus for CPGE and GME as well) are given in Table \ref{table:symmetry}. 

\emph{Gyrotropic point groups} -- The term `gyrotropic' can generate some confusion in the literature since some works use it interchangeably with `optically active', but others distinguish optically active from `weakly gyrotropic'. 

Prior to the work of Fedorov in 1959~\cite{Fedorov59,Fedorov73} the rotation of the plane of polarization of linearly polarize light was taken as a definition for gyrotropy and was assumed to be the same as having a non-zero gyrotropy tensor $g_{ij}$. These are the classes known as optically active (see OA column in Table~\ref{table:symmetry}) and add up to a total of 15 point groups. However, Fedorov showed~\cite{Eritsyan82} that three more point groups should be called gyrotropic. These are $C_{4v}$, $C_{3v}$ and $C_{6v}$, marked wG for `weakly gyrotropic' in the table. The reason to call them gyrotropic is that the gyrotropy tensor is also non-zero. However, since in these three classes this tensor is fully antisymmetric it does not rotate the plane of polarization. As described in Appendix~\ref{appendix:GME}, the rotatory power is determined only by the symmetric part of the gyrotropy tensor~\cite{AG84,AG62,Nye}. 

Weakly gyrotropic crystals differ from non-active crystals in the sense that light reflected from them is elliptically polarized~\cite{Priou97}. Thus, the number of crystallographic classes in which gyrotropy is possible is 18 (marked under G in Table~\ref{table:symmetry}), and not only the optically active 15. 

To summarize, the antisymmetric part of the gyration tensor does not enter the rotation of the polarization plane of a transmitted wave. Therefore, when a material has a zero symmetric part and a non-zero antisymmetric part, the material is called weakly gyrotropic and elliptically polarizes a reflected wave. When the symmetric part of the gyration tensor is non-zero, irrespective of the antisymmetric part, the material is referred to as optically active because it rotates the plane of polarization. Both of these together are the gyrotropic point groups. We also note that some authors use the terms `optically active' and `gyrotropic' interchangeably, but we prefer to distinguish them as explained here.

\emph{Piezoelectric point groups} -- Piezoelectricity is a current response to an applied mechanical strain $u_{ij}$. There are 20 piezoelectric point groups~\cite{BS80,SturmanBook} which are all the non-centrosymmetric ones except the O group. 

\emph{Point groups with finite Linear and Circular Photogalvanic effects} -- The linear photogalvanic effect is a current response to a symmetric tensor $E_iE_j^*+E_i^*E_j$ and the symmetry constraints are therefore the same as for piezoelectricity. As stated in the main text, the circular photogalvanic effect is the current response to a pseudovector $\mathbf{E}\times \mathbf{E}^*$ and therefore it has the same symmetry constraints as gyrotropy~\cite{BS80}. 

Finally, the diagonal elements of the CPGE and GME tensors are zero for all non-enantiomorphic point groups except $S_4$. This point group has no mirrors, so the response tensors have non-vanishing elements in the diagonal; but their sum (the trace) is zero due to the improper rotation. Therefore only enantiomporphic point groups can show a nonzero quantized trace of CPGE, or a  GME.

%
\section{Analytical Solution of $H_{3f}$ \label{appendix:H3f}}
%

In this appendix we derive the analytical eigenstates and energies of $H_{3f}$ Eq.~\eqref{eq:H3f} for general $\phi$, and in particular we show explicitly that the quantity $R^i_{nm}$ is radial. The energies can be obtained from
\begin{equation}\label{eq:cubic}
{\rm det} (H - \mathcal{I} E) = - E^3 + E k^2 + 2 k_x k_y k_z \cos 3 \phi =0
\end{equation}
This is a cubic equation without quadratic term (known as depressed cubic). The three solutions can be written in closed form as
\begin{equation}\label{eq:energies}
E_n = \tfrac{2|k| }{\sqrt{3}} \cos \left(\tfrac{1}{3} \arccos \left(\tfrac{3\sqrt{3} k_x k_y k_z}{k^3} \cos 3\phi \right) - \tfrac{2\pi (n-1)}{3}\right)
\end{equation}
for $n=1,2,3$, and with $0<\arccos x < \pi$. When $\phi = \pi/2$ we have $E_1 = k$, $E_2 = 0$ and $E_3 = -k$. Note $k_x k_y k_z /k^3$ takes its maximum value in the (1,1,1) direction where $k_x k_y k_z /k^3 = 1/(3\sqrt{3})$. In this direction we have 
\begin{equation}
E_n^{(1,1,1)} = \tfrac{2}{\sqrt{3}} |k| \cos \left(\phi- \tfrac{2\pi (n-1)}{3}\right)
\end{equation}
where due to the definition of $\arccos x$, $\phi$ should be understood in the sector $0 < \phi < \pi/3$. And the splitting between bands is given by $E_{12} = 2|k| \cos (\phi +\pi/6)$, $E_{23} = 2|k| \sin \phi$, $E_{13} = 2|k| \cos (\phi - \pi/6)$. In the $(-1,-1,-1)$ direction the same expressions hold with $\phi \rightarrow -\phi + \pi/3$. 

The manifold $S_{12}$ becomes active for $\omega>\omega_0$ and is closed for $\omega_1 < \omega <\omega_2$ while  $S_{13}$ becomes active for $\omega> \omega_3$ and fully closed for $\omega > \omega_4$. $S_{23}$ becomes active with $\omega> \omega_5$ and is never closed in the linear model. These frequencies are 
\begin{align}\label{eq:CPGE_frequencies}
\omega_0 &= \mu\frac{\sqrt{3} \cos (\phi + \pi/6 )}{ \cos(\phi)} &
\omega_1 &= \mu\frac{\sqrt{3} \cos (\phi + \pi/6 )}{ \cos(\phi - 2\pi/3)} &
\omega_2 &= \mu\frac{\sqrt{3}\cos(-\phi+\pi/2)}{\cos(-\phi+\pi/3)} \nonumber \\
\omega_3  &=\mu\frac{\sqrt{3}\cos(\phi-\pi/6)}{\cos(\phi)} &
\omega_4  &=\mu \frac{\sqrt{3}\cos(-\phi+\pi/6)}{\cos(-\phi+\pi/3)} &
\omega_5  &=\mu \frac{\sqrt{3}\sin \phi}{\cos(\phi-2\pi/3)}
\end{align}

The normalized eigenfunctions that correspond to $E_n$ are 
\begin{align}
\label{eq:3feigenvalues}
\psi_n = \frac{1}{\sqrt{(3E_n^2 -k^2)(E_n^2-k_z^2)}} \left(\begin{array}{c} E_n^2-k_z^2 \\ E_nk_x e^{-i\phi}+k_yk_ze^{2i\phi} \\ E_n 
k_y e^{i\phi}+k_x k_z e^{-2i\phi} \end{array}\right)
\end{align}
To prove that this is indeed an eigenvector, Eq.~\eqref{eq:cubic} was used with $E=E_n$. These wavefunctions can be used to obtain the diagonal velocity matrix elements as
\begin{align}
v^i_{nn} = \left<\psi_n| \partial_{k_i}H | \psi_n \right> & = \frac{2E_n k_i +2(k_x k_y k_z/k_i) \cos 3 \phi}{3 E_n^2-k^2}
\end{align}
For the CPGE integral the difference $\Delta^i_{nm} = v^i_{nn} - v^i_{mm}$ is needed and is given by
\begin{align}
&\Delta^i_{nm} =  \frac{2E_{mn}\left[k_i(3E_nE_m+k^2)+3(E_m+E_n)\tfrac{k_xk_yk_z}{k_i}\cos 3\phi\right]}{(3E_n^2-k^2)(3E_m^2-k^2)} 
\end{align}
The quantities $R_{nm}^i = r^j_{nm}r^k_{mn}\epsilon^{ijk} = \tfrac{1}{(E_n-E_m)^2} \left<n|\partial_jH|m\right> \left<m|\partial_kH|n\right>\epsilon^{ijk}$ are given by
\begin{align}
R_{nm}^i &= k^i \frac{2 (E_n+E_m)^2\left(E_n^2 E_m^2  +(E_n E_m-k^2)k_z^2+k_z^4\right)\sin 3 \phi}{(E_n-E_m)(E_n^2-k_z^2)(-3 E_n^2 + k^2)(E_m^2-k_z^2)(-3 E_m^2 + k^2)}
\end{align}
Where we have used 
\begin{align}\label{eq:trick1}
E_n^2+E_m^2+E_nE_m = k^2
\end{align}
and 
\begin{align}\label{eq:trick2}
(E_n+E_m)E_nE_m = -2 k_xk_yk_z \cos 3\phi
\end{align}
which can be obtained by subtracting Eq.~\eqref{eq:cubic} for $E_n$ and $E_m$ provided $n\neq m$. This shows that $R_{nm}^i$ is indeed purely radial. 

%
\section{Energy Scales for $H_{4f}$}\label{app:4f}
%

The eigenvalues of $H_{4f}$ in Eq.~\eqref{eq:4f} can be obtained from 
\begin{equation}
{\rm det} (H_{4f} - E \mathcal{I}) = E^4 - E^2 \mathbf{k}^2 + f(\mathbf{k},\chi) =0
\end{equation}
with 
\begin{equation}
f(\mathbf{k},\chi) = \tfrac{1}{8}(1-\cos 4\chi)(k_x^4+k_y^4+k_z^4)+\tfrac{1}{8}(\tfrac{11}{4}+\tfrac{7}{4}\cos 4\chi +3 \sin 2\chi)(k_x^2k_y^2 + k_x^2 k_z^2 +k_y^2k_z^2).
\end{equation}
The solutions in decreasing order are given by
\begin{align}
E_1(\mathbf{k}) =& \sqrt{\frac{\mathbf{k}^2+\sqrt{\mathbf{k}^4-4f(\mathbf{k},\chi)}}{2}} &
E_2(\mathbf{k}) =& \sqrt{\frac{\mathbf{k}^2-\sqrt{\mathbf{k}^4-4f(\mathbf{k},\chi)}}{2}} \nonumber \\
E_3(\mathbf{k}) =& -\sqrt{\frac{\mathbf{k}^2-\sqrt{\mathbf{k}^4-4f(\mathbf{k},\chi)}}{2}} &
E_4(\mathbf{k}) =& -\sqrt{\frac{\mathbf{k}^2+\sqrt{\mathbf{k}^4-4f(\mathbf{k},\chi)}}{2}}
\end{align}
Defining $\mathbf{k}^{100} = k(1,0,0)$ and $\mathbf{k}^{111} = k(1,1,1)/\sqrt{3}$, the different frequencies defined in the main text are given by
\begin{align}
\omega_0 &= \mu \frac{E_1(\mathbf{k}^{111})-E_2(\mathbf{k}^{111})}{E_1(\mathbf{k}^{111})} & \omega_1 = \mu \frac{E_1(\mathbf{k}^{100})-E_2(\mathbf{k}^{100})}{E_1(\mathbf{k}^{100})} & & \omega_2 = \mu \frac{E_1(\mathbf{k}^{100})-E_2(\mathbf{k}^{100})}{E_2(\mathbf{k}^{100})}   \nonumber \\
\omega_3 &= \mu \frac{E_1(\mathbf{k}^{111})-E_2(\mathbf{k}^{111})}{E_2(\mathbf{k}^{111})}  & \omega_4 = \mu \frac{E_1(\mathbf{k}^{100})-E_3(\mathbf{k}^{100})}{E_1(\mathbf{k}^{100})} & & \omega_5 = \mu \frac{E_1(\mathbf{k}^{111})-E_3(\mathbf{k}^{111})}{E_1(\mathbf{k}^{111})}  \nonumber  \\
\omega_6 &= \mu \frac{E_1(\mathbf{k}^{111})-E_4(\mathbf{k}^{111})}{E_1(\mathbf{k}^{111})}  & \omega_7 = \mu \frac{E_2(\mathbf{k}^{111})-E_4(\mathbf{k}^{111})}{E_2(\mathbf{k}^{111})} & & \omega_8 = \mu \frac{E_2(\mathbf{k}^{100})-E_4(\mathbf{k}^{100})}{E_2(\mathbf{k}^{100})}   
\end{align}

%
\section{\emph{Ab Initio} Calculation Methods}\label{appendix:ab_initio}
%

Calculations have been performed within the DFT\cite{Hohenberg-PR64,Kohn-PR65} as implemented in the Vienna Ab initio Simulation Package (VASP)\cite{Kresse199615,PhysRevB.48.13115}. The interaction between ion cores and valence electrons was treated by the projector augmented-wave method\cite{vaspPaw}, the generalized gradient approximation (GGA) for the exchange-correlation potential with the Perdew-Burke-Ernkzerhof for solids parametrization~\cite{PBE} and spin-orbit coupling was taken into account by the second variation method\cite{PhysRevB.62.11556}. A Monkhorst-Pack
k-point grid of (4$\times$4$\times$4) for reciprocal space integration and 500 eV energy cutoff of the plane-wave expansion have been used to get a residual error on the energy of less than 10$^{-3}$ meV, resulting in a fully converged electronic structure including spin-orbit coupling.

%
\section{Tight-binding Band Structures}\label{appendix:tight_binding}
%

In this appendix, we review our construction of tight-binding models in space groups 199, 214, and 198, paying particular attention to the boundary conditions imposed by the atomic positions. Subsection~\ref{subsec:199} reviews the construction of the nearest-neighbor model for space groups 199 and 214 without spin-orbit coupling, largely following the discussion in Ref.~\onlinecite{manes2012}. In Subsection~\ref{subsec:198} we show how to modify the tight binding model of Ref.~\onlinecite{ChangEA17} to accurately describe the atomic positions of RhSi.

\subsection{SGs 199 and 214}\label{subsec:199}

Here we will construct minimal tight-binding models for space groups 199 and 214 without spin-orbit coupling. Both of these groups are body-centered cubic, with Bravais lattice vectors
\begin{align}
R_1&=\frac{a}{2}(-\mathbf{\hat{x}}+\mathbf{\hat{y}}+\mathbf{\hat{z}})\nonumber\\
R_2&=\frac{a}{2}(\mathbf{\hat{x}}-\mathbf{\hat{y}}+\mathbf{\hat{z}})\nonumber\\
R_3&=\frac{a}{2}(\mathbf{\hat{x}}+\mathbf{\hat{y}}-\mathbf{\hat{z}}).
\end{align}
For simplicity, we will take the lattice constant $a=1$ for the remainder of this work. Note first that space group 199 is generated by
\begin{equation}
G_{199}=\langle\{C_{2x}|\half\half 0\},\{C_{3,111}|000\},\{E|100\},\{E|010\},\{E|001\}\rangle
\end{equation}
Here $E$ denotes the identity rotation, and the translation part of space group elements will be given in reduced coordinates, \emph{i.e.}
\begin{equation}
\{E|\half\half 0\}\rightarrow \half R_1 + \half R_2.
\end{equation}
Space group 214 is obtained by appending to this generating set the additional twofold screw
\begin{equation}
G_{214}=\langle G_{199}, \{C_{2,110}|\half 0 0\}\rangle.
\end{equation}

To construct a minimal tight binding model, we will place spinless $s$-orbitals at the minimal-multiplicity Wyckoff position in the space group, and consider nearest neighbor hoppings. In both SGs 199 and 214, the minimal-multiplicity Wyckoff position is the $8a$ position (and also the $8b$ position in SG 214), with multiplicity 4 in both cases. Because the stabilizer group of this position in SG214 contains $C_{2,110}$, the tight binding models for both SG199 and 214 will be formally identical. Concretely, the four points in the $8a$ (or $b$) position are, in reduced coordinates
\begin{equation}
\{\mathbf{q}_1,\mathbf{q}_2,\mathbf{q}_3,\mathbf{q}_4\}=\{(u,u,u),(\half-u,\half,0),(0,\half-u,\half),(\half,0,\half-u)\},
\end{equation}
obtained by acting successively with $\hat{x},\hat{y}$ and $\hat{z}$ twofold rotations on $\mathbf{q}_1$. For SG199 the value of $u$ is arbitrary, $-\half<u<\half$, while for SG 214, $u$ is fixed to $\pm \frac{1}{4}$. As this will be the only difference between the two models, we will leave $u$ arbitrary. For simplicity, we will take $u>0$ without loss of generality. Our final model will be applicable for either sign of $u$.

By using $s$-orbitals as our basis functions, all symmetries in the stabilizer group of the $8a$ position are represented trivially. Thus, to construct our model, we need only ensure that all hoppings appear in symmetry invariant combinations. Note that the shortest distance between orbitals in the lattice is given by
\begin{equation}
|\mathbf{q}_1-\mathbf{q}_j|^2=\frac{1}{4}-u+2u^2 < \frac{1}{4},
\end{equation}
equal for all all symmetry related bonds. Thus, we find for the nearest neighbor Hamiltonian
\begin{equation}
H_{NN}=t_{NN}\sum_{\mathbf{R}}\left(c^\dag_{2,\mathbf{R}}c^{\vphantom{\dag}}_{1,\mathbf{R}}+c^\dag_{3,\mathbf{R}}c^{\vphantom{\dag}}_{1,\mathbf{R}}+c^\dag_{4,\mathbf{R}}c^{\vphantom{\dag}}_{1,\mathbf{R}}+c^\dag_{4,\mathbf{R}-\mathbf{R}_1}c^{\vphantom{\dag}}_{3,\mathbf{R}}+c^\dag_{4,\mathbf{R}+\mathbf{R}_2}c^{\vphantom{\dag}}_{2,\mathbf{R}}+c^\dag_{3,\mathbf{R}-\mathbf{R}_3}c^{\vphantom{\dag}}_{2,\mathbf{R}}\right)+\mathrm{h.c.}
\end{equation}
where $c_{i,\mathbf{R}}$ annihilates a fermion at site $i$ in unit cell $\mathbf{R}$. We can Fourier transform this using
\begin{equation}
c_{i,\mathbf{R}}=\sum_k e^{i,\mathbf{k}\cdot(\mathbf{R}+\mathbf{q}_i)}c_{i\mathbf{k}}
\end{equation}
to obtain
\begin{equation}
H_{NN}=t_{NN}\sum_\mathbf{k}
\left(\begin{array}{cccc}
c^\dag_{1,\mathbf{k}} & c^\dag_{2,\mathbf{k}} & c^\dag_{3,\mathbf{k}} & c^\dag_{4,\mathbf{k}}
\end{array}\right)
V^\dag(u,\mathbf{k})H_0(\mathbf{k})V(u,\mathbf{k})
\left(\begin{array}{c}
c^\dag_{1,\mathbf{k}} \\
 c^\dag_{2,\mathbf{k}} \\
 c^\dag_{3,\mathbf{k}} \\
 c^\dag_{4,\mathbf{k}}
\end{array}
\right)
\end{equation}
with
\begin{equation}
H_0(\mathbf{k})=\left(\begin{array}{cccc}
0 & 1 & 1 & 1 \\
1 & 0 & e^{-i\mathbf{k}\cdot\mathbf{R}_3} & e^{i\mathbf{k}\cdot\mathbf{R}_2} \\
1 & e^{i\mathbf{k}\cdot\mathbf{R}_3} & 0 & e^{-i\mathbf{k}\cdot\mathbf{R}_1}\\
1 & e^{-i\mathbf{k}\cdot\mathbf{R}_2} & e^{i\mathbf{k}\cdot\mathbf{R}_1} & 0
\end{array}\right)
\end{equation}
and
\begin{equation}
V(u,\mathbf{k})=\left(\begin{array}{cccc}
e^{i\mathbf{k}\cdot\mathbf{q}_1} & 0 & 0 & 0 \\
0 & e^{i\mathbf{k}\cdot\mathbf{q}_2} & 0 & 0 \\
0 & 0 & e^{i\mathbf{k}\cdot\mathbf{q}_3} & 0 \\
0 & 0 & 0 & e^{i\mathbf{k}\cdot\mathbf{q}_4}
\end{array}
\right).
\end{equation}
Note that only the matrix $V$, which determines the embedding of the orbitals, distinguishes between SG 199 and 214 in this model.

\subsection{RhSi and SG 198}\label{subsec:198}

\begin{figure}  
\includegraphics[width=0.5\linewidth]{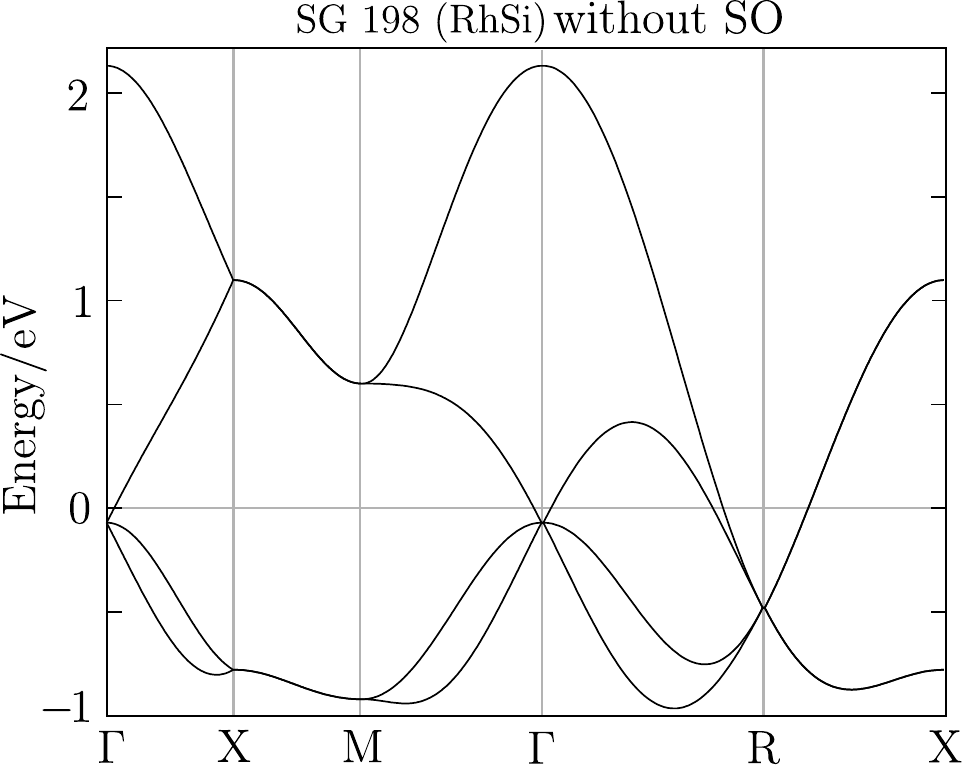}
\caption{
{\bf RhSi bandstructure without spin-orbit coupling}. The real material features a small spin-orbit coupling as shown in Fig.~\ref{fig:TBbands} in the main text.
}
\end{figure}

In this section, we introduce some modifications to the tight-binding model of RhSi given in Ref.~\onlinecite{ChangEA17}, in order to obtain more physically meaningful results for the GME and CPGE. In particular, we focus on the embedding and boundary conditions on Bloch functions. The tight-binding model without spin-orbit coupling is shown in Fig.~\ref{appendix:tight_binding}1, and that with spin-orbit coupling was shown in Fig.~\ref{fig:TBbands}. First, we note that Ref.~\onlinecite{ChangEA17} chose for the locations of their atoms
\begin{align}
\mathbf{q}_A&=(0,0,0)\nonumber\\
\mathbf{q}_B&=(\half,\half,0)\nonumber\\
\mathbf{q}_C&=(\half,0,\half)\nonumber\\
\mathbf{q}_D&=(0,\half,\half),
\end{align}
given in reduced coordinates (which here {are aligned with the cartesian directions}). Introducing the matrix
\begin{equation}
V(\mathbf{k})=\exp\left[\left(
\begin{array}{cccc}
0 & 0 & 0 & 0 \\
0 & \frac{i}{2}(k_1+k_2) & 0 & 0 \\
0 & 0 & \frac{i}{2}(k_1+k_3) & 0 \\
0 & 0 & 0 & \frac{i}{2}(k_2+k_3)
\end{array}
\right)\right]\otimes\sigma_0
\end{equation}
we have that their tight-binding Hamiltonian $\mathcal{H}(\mathbf{k})$ satisfies
\begin{equation}
\mathcal{H}(\mathbf{k}+\mathbf{G})=V(\mathbf{G})^\dag \mathcal{H}(\mathbf{k}) V(\mathbf{G}).
\end{equation}
The rows and columns of our matrices correspond to orbitals $A,B,C$ and $D$, in that order. 

However, these are not boundary conditions and atomic coordinates of the dominant states near the Fermi energy in RhSi. Our ab-initio calculations reveal that the relevant orbitals near $E_F$ originate from Rh atoms in this material, which are located at 
\begin{align}
\mathbf{q}_A&=(x,x,x)\nonumber\\
\mathbf{q}_B&=(\half+x,\half-x,-x)\nonumber\\
\mathbf{q}_C&=(\half-x,-x,\half+x)\nonumber\\
\mathbf{q}_D&=(-x,\half+x,\half-x),
\end{align}
with $x=0.3959$. To obtain a tight-binding Hamiltonian in the proper embedding, we must take
\begin{equation}
\mathcal{H}_x(\mathbf{k})\equiv U_x(\mathbf{k})^\dag \mathcal{H}(\mathbf{k})U_x(\mathbf{k})
\end{equation}
with
\begin{equation}
U_x(\mathbf{k})=\exp\left[\left(
\begin{array}{cccc}
ix(k_1+k_2+k_3) & 0 & 0 & 0 \\
0 & ix(k_1-k_2-k_3) & 0 & 0 \\
0 & 0 & ix(k_3-k_2-k_1) & 0 \\
0 & 0 & 0 & ix(k_2-k_1-k_3)
\end{array}
\right)\right]\otimes\sigma_0
\end{equation}
With this choice, we can easily verify that 
\begin{equation}
\mathcal{H}_x(\mathbf{k}+\mathbf{G})_{\alpha\beta}=e^{-i\mathbf{G}\cdot\mathbf{q}_\alpha}\mathcal{H}_x(\mathbf{k})_{\alpha\beta} e^{i\mathbf{G}\cdot\mathbf{q}_\beta},
\end{equation}
where there is no implied summation over repeated indices. 

%
\section{Low energy Hamiltonians for the Double Spin-1/2 and Double Spin-1 Fermions}\label{appendix:kdphams}
%

In this section, we elaborate on some details of the derivation of the $\mathbf{k}\cdot\mathbf{p}$ Hamiltonians for the double spin-1/2 and double spin-$1$ fermions given respectively in Eqs.~(\ref{eq:4f}) and (\ref{eq:H6f}) in the text. We also derive the low energy Hamiltonian for tetrahedral spin-$3/2$ fermions. In this Appendix, $k_i$ will be used to denote the displacement away from a high-symmetry momentum $\mathbf{K}$ for convenience.

\subsection{Double spin-$1/2$ Fermions}

As mentioned in Sec.~\ref{sec:multifold_intro}, double spin-1/2 fermions are protected by the combination of a perpendicular twofold screw rotational symmetry and time-reversal symmetry. Due to the chiral nature of the space groups in question, this fourfold degeneracy takes the form of two Weyl points \emph{of the same charge} pinned to lie at the same energy, with symmetry-allowed inter-node coupling. 

Following the methods of Ref.~\onlinecite{BradlynEA17}, the low energy $\mathbf{k}\cdot\mathbf{p}$ Hamiltonian for the double spin-1/2 fermions in SG 90 can be shown to be 
\begin{equation}\label{eq:h90coupled}
H_{90}(\mathbf{k})=\left(
\begin{array}{cccc}
a k_z & b \epsilon k_+ & ck_- & 0 \\
b \epsilon^*k_- & -a k_z & 0 & i c k_+ \\
c^*k_+ & 0 & -a k_z & -b \epsilon^* k_-\\
0 & -i c^*k_-& -b  \epsilon k_+ & a  k_z
\end{array}
\right),
\end{equation}
where $\epsilon = e^{-i\pi/4}$, $a$ and $b$ are real parameters, and $c$ is complex. This Hamiltonian is written in the basis where the time reversal operator is the natural one, 
\begin{equation}
\rho(T) = \left(\begin{array}{cc}
0 & \mathbb{I} \\
-\mathbb{I} & 0
\end{array}\right) \mathcal{K},
\end{equation}
where $\mathcal{K}$ is complex conjugation; and the spatial symmetries are represented by the block-diagonal matrices
\begin{equation}
\rho(C_{4z})=\left(\begin{array}{cccc}
e^{-3\pi i/4} & 0 & 0 & 0 \\
0 & e^{-i\pi/4} & 0 & 0 \\
0 & 0 & e^{3\pi i /4} & 0 \\
0 & 0 & 0 & e^{i\pi/4}
\end{array}
\right),\;\;
\rho(\{C_{2x}|\half\half 0\})=\left(\begin{array}{cccc}
0 & e^{-i\pi/4} & 0 & 0 \\
e^{i\pi/4} & 0 & 0 & 0 \\
0 & 0 & 0 & e^{i\pi/4} \\
0 & 0 & e^{-i\pi/4} & 0
\end{array}
\right)
\end{equation}

Note that even though there is coupling between the Weyl fermions, along the plane $\delta k_x=0$, bands remain doubly degenerate. This is a generic feature of these double spin-$1/2$ fermions, and is due to the fact that they are protected by a twofold screw symmetry $g=\{C_{2x}|\frac{1}{2}\frac{1}{2}0\}$. Because the double spin-$1/2$ occurs at a point $\mathbf{K}$ with $K_x=1/2$, we have that the product of $g$ and time-reversal symmetry leaves the plane $k_x=0$ invariant and squares to $-1$, thus enforcing a Kramers degeneracy.

By applying a constant unitary rotation to Eq.~(\ref{eq:h90coupled}), we can transform it to a basis where the two Weyl fermions are decoupled. First, writing $c=|c|e^{i\phi}$ we note that the operator
\begin{equation}
A=\left(\begin{array}{cccc}
0 & 0 & 0 & -e^{i(\phi-\pi/4)} \\
0 & 0 & e^{i(\phi-\pi/4)} & 0 \\
0 & e^{-i(\phi-\pi/4)} & 0 & 0 \\
-e^{-i(\phi-\pi/4)} & 0 & 0 & 0
\end{array}\right)
\end{equation}
commutes with Eq.~\eqref{eq:h90coupled}, and has eigenvalues $(1,1,-1,-1)$. Re-expressing the Hamiltonian in the eigenbasis of $A$ and reabsorbing the constant phase $\epsilon$ into the definition of the basis states we recover Eq.~\eqref{eq:H90} of the main text. 

The double spin-1/2 fermion in space group 198 at the $M$ point has a similar, but slightly less constrained Hamiltonian, due to the absence of fourfold rotational symmetry. The little group at the $M$ point has a four dimensional physically irreducible representation which can be expressed as
\begin{equation}
\rho(\{C_{2x}|\half\half 0\})=\sigma_z\tau_0,\;\; \rho(\{C_{2z}\half 0\half\})=i\sigma_y\tau_0,\;\; \rho(T)=-i\sigma_0\tau_y\mathcal{K},
\end{equation}
which lead to the linear Hamiltonian
\begin{equation}
H_{198}=ck_z\sigma_y\tau_0+k_x\sigma_z\vec{v}_1\cdot\vec{\tau}+k_y\sigma_x\vec{v}_2\cdot\vec{\tau},
\end{equation}
where $\vec{\sigma}$ and $\vec{\tau}$ are vectors of Pauli matrices, $\tau_0$ is the identity matrix in ``$\tau$-space'', and the tensor product of $\sigma$ and $\tau$ is implied. Additionally, $c$ is a real scalar, and $\vec{v}_1,\vec{v}_2$ are three-vectors of real parameters. The additional degrees of freedom in this Hamiltonian compared to the one in SG 90 arise due to the absence of $C_4$ symmetry. Nevertheless, we can still decouple this double spin-1/2, by noting that the operator 
\begin{equation}
A=\frac{1}{|\vec{v}_1\times\vec{v}_2|}\sigma_y(\vec{v}_1\times\vec{v}_2)\cdot\vec{\tau}
\end{equation}
commutes with $H_{198}$, and has eigenvalues $(1,1,-1,-1)$. Expressing the Hamiltonian in terms of the eigenspaces of $A$ thus generically decouples the Hamiltonian. The dispersion of this Hamiltonian is quite complicated, and takes the form
\begin{equation}
\epsilon_{\pm\pm}=\pm\sqrt{|\vec{v}_1k_x|^2+|\vec{v}_2k_y|^2\pm2|\vec{v}_1\times\vec{v}_2|k_xk_y+c^2k_z^2}.
\end{equation}
Finally, note that when $\vec{v}_1\parallel\vec{v}_2$ the operator $A$ is not defined, however in this case the Hamiltonian is trivially decoupled by $A'=\hat{v}_1\cdot\vec{\tau}$.
\subsection{Double spin-$1$ Fermion}

The low energy Hamiltonian for the doubled $S=1$ fermion was given in Ref.~\onlinecite{BradlynEA17} as
\begin{align}\label{eq:6f}
H_{6f}\left(\boldsymbol{k}\right)=\left(\begin{array}{cc}
H_{3f}\left(\phi,\boldsymbol{k}\right) & bH_{3f}\left(0,\boldsymbol{k}\right)\\
b^{*}H_{3f}\left(0,\boldsymbol{k}\right) & -H_{3f}^{*}\left(\phi,\boldsymbol{k}\right)
\end{array}\right).
\end{align}
Noting that $H_{6f}$ commutes with the operator
\begin{align}
A&=
\begin{pmatrix}
\cos \phi & b \\
b^* & -\cos \phi \\
\end{pmatrix}
\otimes\mathcal{I}_{3 \times 3},
\end{align}
$H_{6f}$ can be decoupled into two blocks labeled by the eigenvalues of $A$, $\pm \sqrt{\cos^2 \phi + |b|^2}$. The result takes the form
\begin{align}\label{eq:H6fapp}
H_\mathrm{6f} &= \sqrt{1+|b|^2}
\left(\begin{array}{cc}
H_{3f}(\tfrac{\pi}{2}-\delta \phi,\boldsymbol{k}) & 0\\
0 &H_{3f}(\tfrac{\pi}{2}+\delta \phi,\boldsymbol{k})
\end{array}\right)
\end{align}
with $\delta \phi = \tan^{-1} \left(\sqrt{\cos^2 \phi + |b|^2} /\sin \phi\right)$. For $H_\t{212,213} \equiv H_\t{198}(\phi=\pi/2)$ this reduces to $\delta \phi = \tan^{-1} |b|$. 
\subsection{Tetrahedral Spin-$3/2$ Fermions}
The tetrahedral spin-$3/2$ fermions in space groups 195--198 arise from the octahedral spin-$3/2$ fermions in space groups 207--214 upon the breaking of fourfold rotational symmetry. These relaxed symmetry constraints allow for an additional term in the $\mathbf{k}\cdot\mathbf{p}$ Hamiltonian at linear order, which takes the form
\begin{equation}
H_{4f,T}=H_{4f}+v_T\left(
\begin{array}{cccc}
0 & k_z & -\sqrt{3}k_x & ik_y \\
k_z & \frac{2k_z}{\sqrt{3}} & ik_y &\frac{k_x-2ik_y}{\sqrt{3}} \\
-\sqrt{3}k_x & -ik_y & 0 & -k_z \\
-ik_y & \frac{k_x+2ik_y}{\sqrt{3}} & -k_z & -\frac{2k_z}{\sqrt{3}}
\end{array}
\right),
\end{equation}
where $H_{4f}$ is the octahedral spin-$3/2$ Hamiltonian given in Eq.~(\ref{eq:4f}), and $v_T$ is a real parameter proportional to the strength of $C_4$ symmetry breaking. This gives the most general linear Hamiltonian invariant under the $^{1}\bar{F}^{2}\bar{F}$ (co-)representation of the tetrahedral group, which is generated by
\begin{align}
\rho(C_{2x})=-i\left(\begin{array}{cccc}
0 & 0 & 1 & 0 \\
0 & 0 & 0 & 1 \\
1 & 0 & 0 & 0 \\
0 & 1 & 0 & 0
\end{array}
\right),\;\;&
\rho(C_{3,(1,-1,1)})=\frac{1}{4}\left(\begin{array}{cccc}
1-i & \sqrt{3}(1-i) & 1+i & (1+i)\sqrt{3} \\
(-1+i)\sqrt{3} & 1-i & -\sqrt{3}(1+i) & 1+i \\
-1+i & \sqrt{3}(-1+i) & 1+i & \sqrt{3}(1+i) \\
\sqrt{3}(1-i) & -1+i & -\sqrt{3}(1+i) & 1+i
\end{array}
\right),\\
\rho(T)&=\left(\begin{array}{cccc}
0 & 0 & -1 & 0 \\
0 & 0 & 0 & -1 \\
1 & 0 & 0 & 0 \\
0 & 1 & 0 & 0
\end{array}
\right)\mathcal{K},
\end{align}
\end{widetext}

\end{document}